\documentclass[letterpaper,11pt,leqno]{article}
\usepackage[T1]{fontenc}
\usepackage{newtxtext}
\RequirePackage[scaled=.9]{helvet}
\usepackage[dvipsnames,svgnames,x11names]{xcolor}
\usepackage{orcidlink}

\usepackage[authoryear]{natbib}     
\usepackage{hyperref}
\hypersetup{bookmarksnumbered=true, colorlinks=true,
    citecolor=ForestGreen,       
    linkcolor=Blue, 		     
    urlcolor=MidnightBlue     	 
}

\usepackage{amsthm,amsmath,amsfonts,amssymb}
\usepackage{physics,mathrsfs,bm,bbm}
\usepackage{enumitem,multicol,multirow}
\usepackage{graphicx,subfig,tikz}
\graphicspath{{figure/}{art/}}

\usepackage[linesnumbered,ruled]{algorithm2e}
\SetKwRepeat{Do}{do}{while}

\theoremstyle{plain}
\newtheorem{theorem}{Theorem}
\newtheorem{lemma}{Lemma}

\theoremstyle{remark}

\newtheorem{example}{Example}
\theoremstyle{definition}

\usepackage{etoolbox}
\newcommand{\renewSformat}[1]{%
  \expandafter\renewcommand\csname the#1\endcsname{S\arabic{#1}}%
  \setcounter{#1}{0}%
}

\newcommand{\R}{\mathbb{R}}					
\newcommand{\bN}{\mathbb{N}}				
\newcommand{\N}{\mathsf{N}}		       		
\newcommand{\1}{\mathbbm{1}}				
\renewcommand{\Pr}{\mathbb{P}}		
\newcommand{\E}{\mathbb{E}}			
\newcommand{\Var}{\operatorname{Var}}		
\newcommand{\normvec}[2][]{#1\lVert #2 #1\rVert_{\infty}}
\newcommand{\normsum}[2][]{#1\lVert #2 #1\rVert_{1}}
\newcommand{\normop}[2][]{#1\lvert\!#1\lvert\!#1\lvert #2 #1\rvert\!#1\rvert\!#1\rvert_{\infty}}

\usepackage{authblk}
\title{Algorithm-aware Statistical Inference in High-dimensional Estimating Equations}
\author{Tongyu Li\orcidlink{0009-0007-5051-2716}}
\author{Alexander Giessing\orcidlink{0000-0002-6917-0652}}
\affil{Department of Statistics and Data Science, National University of Singapore\\ \texttt{tongyuli@nus.edu.sg, giessing@nus.edu.sg}}
\date{}

\begin{document}
\maketitle

\begin{abstract}
Iterative algorithms are integral to many statistical procedures, whereas sampling uncertainty along their iteration is generally not quantified. 
We develop simultaneous inference for discrete trajectories generated from high-dimensional estimating equations, with iteration-specific population estimands that preserve algorithmic regularization.
Our proposed fixed-Jacobian bootstrap propagates perturbations along the observed trajectory without repeatedly running the original nonlinear algorithm.
We establish uniform first-order expansions and bootstrap validity over possibly infinite sets of contrasts and iterations under stability, entropy and moment conditions.
Simulations in sparse linear regression, paired Fisher discrimination, and temporal-difference learning demonstrate accurate simultaneous coverage. 
A micro-randomized trial application illustrates inference for a treatment trajectory with participant-level clustering.
\end{abstract}


\section{Introduction}
\label{sec:intro}

\subsection{Background}
Iterative algorithms are essential components of modern statistical methodology \citep{lange2014brief}. 
In high-dimensional problems, a reported estimator is commonly determined not only by the empirical criterion or estimating equation, but also by the algorithm used to compute it.
The algorithmic choices, including initialization, step-size schedule, momentum scheme, thresholding parameters, and stopping rule, can materially affect the estimator’s statistical properties. 
For example, early stopping can regularize unstable or weakly identified directions \citep{hu2022early}, while iterative thresholding can promote sparse intermediate iterates \citep{beck2009fast}. 
More generally, the algorithmic iteration determines the effective spectral filtering of sampling noise \citep{ali2019continuous,celisse2021analyzing}. 
A finite-step iterate and the fully converged optimizer may thus represent distinct statistical procedures. Inference based on the converged optimizer can be transferred to a finite iterate only after showing that their discrepancy is negligible at the relevant inferential scale.

The statistical role of finite iterations has been partly recognized in existing theory. 
For regularized high-dimensional M-estimators, \citet{agarwal2012fast} established geometric convergence of first-order methods up to the statistical precision of the model, beyond which further optimization need not improve statistical accuracy. 
In nonparametric regression over reproducing kernel Hilbert spaces, suitably stopped gradient descent acts as an implicit regularizer and can attain optimal estimation rates \citep{raskutti2014early}. 
These results explain why an intermediate iterate may be statistically meaningful, but they primarily concern estimation or prediction risk rather than joint distributional approximation or simultaneous uncertainty quantification along the algorithmic trajectory.

A growing literature has begun to address inference for algorithmic outputs.
For stochastic-gradient methods, confidence intervals and hypothesis tests have been constructed using covariance estimates based on averaged iterates, online perturbations, and hierarchical branching schemes \citep{chen2020statistical,fang2018online,su2023higrad}.
In high-dimensional linear models, \citet{bellec2026uncertainty} developed estimators of the generalization error along the trajectory, together with debiasing corrections and confidence intervals for the underlying regression coefficients. 
\citet{han2026gradient} obtained iteration-wise debiased inference for gradient descent under proportional asymptotics, while \citet{li2026gradientflows} established time-uniform inference for continuous gradient flows. 
These works demonstrate that optimization dynamics can be incorporated into valid statistical inference.
What remains less developed is simultaneous inference over both contrast directions and iteration indices for general discrete-time recursive procedures.

In this paper, we focus on algorithm- and iteration-specific estimands $\beta_t$ arising from high-dimensional estimating equations. 
The inferential objective is to quantify the uncertainty of $u^\top\hat\beta_t$ for algorithmic outputs $\hat\beta_t$ simultaneously over contrast directions $u$ and iteration indices $t$.
This extends static high-dimensional inference, where debiasing and orthogonalization methods provide confidence intervals for components or linear functionals of a regression or moment parameter \citep{dezeure2015high,belloni2018uniformly}, and multiplier bootstraps yield simultaneous inference over many coordinates \citep{chernozhukov2013gaussian,chernozhukov2023high}. 
For an algorithmic trajectory, the same empirical inputs drive every update, so a common sampling perturbation is propagated recursively through the local dynamics of the algorithm.
Therefore, valid simultaneous inference must capture both the high-dimensional dependence across contrasts and the temporal dependence induced by the iterative recursion.

\subsection{Contributions}
We formulate algorithm-aware statistical analysis for discrete trajectories generated from high-dimensional estimating equations (Section~\ref{sec:framework}). 
The framework permits algorithmic update mappings to depend on the full history of preceding iterates and residuals. Its main example is inertial thresholding, which combines a residual-based update, momentum and coordinatewise soft thresholding. 
The population iterate $\beta_t$ defines the statistical target at iteration $t$, so early stopping and algorithmic regularization are incorporated into the estimand rather than treated as numerical approximation errors.

We propose an algorithm-aware bootstrap that propagates empirical perturbations of the estimating-equation inputs through Jacobian matrices evaluated along the observed trajectory (Section~\ref{sec:inference}). 
These Jacobian matrices remain fixed within each bootstrap replication. Thus, the procedure reproduces the first-order dependence across iterations without rerunning the original nonlinear algorithm on every perturbed input. 

In theoretical analysis, we derive recursive trajectory expansions in which the sampling or bootstrap perturbation is propagated by algorithm-specific sensitivity matrices (Section~\ref{sec:expansion}). 
A finite-sample remainder bound identifies the stability coefficients, local Jacobian continuity and support restrictions that control accumulation across iterations. 
The analysis covers a fixed iteration range and, under uniform stability conditions, an unbounded discrete horizon.
By leveraging Gaussian approximation, anti-concentration and Gaussian comparison, we demonstrate bootstrap validity for maximal statistics indexed jointly by contrasts and iterations (Section~\ref{sec:Gauss}).

The numerical studies exhibit the practical performance of our proposed method (Section~\ref{sec:numeric}).
Simulation covers sparse linear regression, paired Fisher discrimination and least-squares temporal-difference learning, thereby validating distributional approximation in various settings. 
A participant-clustered analysis of the Drink Less micro-randomized trial illustrates simultaneous inference for a high-dimensional treatment trajectory in real world.

\subsection{Notation}
Let $\bN=\{1,2,\dots\}$ be the set of positive integers. 
Write $\norm{\cdot}_q$ for the vector $\ell^q$ norm, $q\in[1,\infty]$, and $\normop{\cdot}$ for the induced $\ell^\infty$ matrix norm (maximum absolute row sum). 
Denote by $\operatorname{vec}$ and $\otimes$ the vectorization and the Kronecker product applied to matrices, respectively. 
Let $\1\{\cdot\}$ be the indicator function. 
For any $v=(v_1,\dots,v_p)^\top\in\R^p$, let $\operatorname{supp}(v)=\{j:v_j\ne0\}$ be its support. 
The shorthands $\Pr_{|W_1^n}$ and $\E_{|W_1^n}$ stand for the conditional probability and expectation given data $W_1,\dots,W_n$. 
For two sequences $(X_n)$ and $(Y_n)$ of nonnegative random variables, $X_n = o_{\Pr}(Y_n)$ means that $\lim_{\epsilon\searrow0} \limsup_{n\to\infty} \Pr(X_n > \epsilon Y_n) = 0$.

\section{Proposed Methodology}
\label{sec:method}

\subsection{Statistical framework and algorithmic target}
\label{sec:framework}
Consider the population estimating mapping 
\[\psi(\beta)=H\beta-\xi,\quad\beta\in\R^p,\]
where $H\in\R^{p\times p}$ and $\xi\in\R^p$ are unknown population quantities. 
The static parameter $\beta^\circ$ is characterized by the population estimating equation $\psi(\beta^\circ)=0_p$.
This formulation accommodates both symmetric and non-symmetric matrices $H$, and thus extends beyond the first-order conditions of quadratic objectives. 
Given the observed data, let $(\hat{H},\hat\xi)$ be a sample analogue of $(H,\xi)$, and define the empirical estimating mapping 
\[\hat\psi(\beta)=\hat{H}\beta-\hat\xi,\quad\beta\in\R^p.\]
In practice, an estimator $\hat\beta$ should be constructed such that the empirical residual $\hat\psi(\hat\beta)$ is small in an appropriate sense. 
Several statistical problems fall within this framework.
\begin{example}[Estimating equations]\label{eg:LEE}
Suppose that $W_1,\dots,W_n$ are independent sampling units.
\begin{enumerate}
\item\emph{Linear regression.}
Consider $W_i=(x_i,y_i)\in\R^p\times\R$. Setting $H=\E(x_ix_i^\top)$ and $\xi=\E(y_ix_i)$, the estimating equation becomes the normal equation.
For clustered least squares, $W_i$ consists of repeated measurements $(x_{ij},y_{ij})\in\R^p\times\R$, $j=1,\dots,m$, whence 
\[H=m^{-1}\sum_{j=1}^{m}\E(x_{ij}x_{ij}^\top) \qand \xi=m^{-1}\sum_{j=1}^{m}\E(y_{ij}x_{ij})\]
correspond to within-cluster averages of the individual contributions.
\item\emph{Paired Fisher discrimination.}
Consider $W_i=(x_i^+,x_i^-)\in\R^p\times\R^p$ where $\E(x_i^\pm)=m^\pm$ and $\Var(x_i^\pm)=H$. Choosing $\xi=m^+-m^-$ leads to the population Fisher direction $\beta^\circ$ for the two classes distinguished by means. 
Along with the usual empirical estimator $\hat\xi=\hat{m}^+-\hat{m}^-$, the matched-pair design results in the pooled within-class covariance estimator 
\[\hat{H}=(\hat{H}^++\hat{H}^-)/2,\quad\hat{H}^\pm=(n-1)^{-1}\sum_{i=1}^{n}(x_i^\pm-\hat{m}^\pm)(x_i^\pm-\hat{m}^\pm)^\top.\]
\item\emph{Least-squares temporal-difference learning \citep{bradtke1996linear,kolter2009regularization}.}
Consider $W_i=(x_i,x_i',y_i)\in\R^p\times\R^p\times\R$ that records state transition and reward. 
To solve Bellman's equation with a discount factor $\delta\in(0,1)$, the parameter $\beta^\circ$ is defined with $H=\E\{x_i(x_i-\delta x_i')^\top\}$ and $\xi=\E(y_ix_i)$. 
In this case, the matrix $H$ is generally non-symmetric.
\end{enumerate}
\end{example}

One may estimate $\beta^\circ$ by directly solving the empirical equation $\hat\psi(\hat\beta)=0_p$ whenever $\hat{H}$ is non-singular and sufficiently well conditioned. 
However, $\hat{H}$ may be singular or unstable to invert, particularly in high-dimensional regimes where $p$ is comparable to or exceeds the sample size. 
Meanwhile, sampling variability generally prevents the empirical quantities from satisfying the population relation exactly. 
These considerations motivate regularized formulations that seek a structured approximate solution to the empirical estimating equation, and practical implementation calls for scalable algorithms that efficiently compute such a solution. 

In this paper, we focus on a broad class of iterative algorithms of the form below. 
Let $F_t:\R^{2tp}\to\R^p$, $t\in\bN$, be deterministic update mappings that allow each update to depend on the entire history of the iterates and their estimating-equation residuals. 
The empirical trajectory is recursively produced as 
\begin{equation}\label{eq:coef_est}
\hat\beta_t=F_t(\hat\theta_{t-1}),\quad
\hat\theta_{t-1}=(\hat\beta_{t-1}^\top,\hat\gamma_{t-1}^\top,\dots,\hat\beta_0^\top,\hat\gamma_0^\top)^\top,
\end{equation}
from a fixed initialization $\hat\beta_0=\beta_0\in\R^p$, where $\hat\gamma_t=\hat\psi(\hat\beta_t)=\hat{H}\hat\beta_t-\hat\xi$ for $t\in\bN\cup\{0\}$. 
Applying the same algorithm to the population estimating equation defines the population trajectory 
\begin{equation}\label{eq:coef_pop}
\beta_t=F_t(\theta_{t-1}),\quad
\theta_{t-1}=(\beta_{t-1}^\top,\gamma_{t-1}^\top,\dots,\beta_0^\top,\gamma_0^\top)^\top,
\end{equation}
where $\gamma_t=\psi(\beta_t)=H\beta_t-\xi$ for $t\in\bN\cup\{0\}$. 
\begin{example}[Iterative algorithm]\label{eg:alg}
Let $\eta_t>0$ be step sizes, $\lambda\ge0$ be a regularization parameter, and $\mu_t\ge0$ be inertial parameters, with $\mu_1=0$ so that the lag-two term is omitted at the first update. 
Consider the update mapping 
\[F_t(\theta_{t-1})=\mathcal{S}_{\lambda\eta_t}\big((1+\mu_t)(\beta_{t-1}-\eta_t\gamma_{t-1})-\mu_t(\beta_{t-2}-\eta_t\gamma_{t-2})\big),\]
where the soft-thresholding $\mathcal{S}_\tau(z)=\operatorname{sign}(z)(\abs{z}-\tau)_+$ acts coordinatewise. 
The resulting algorithm combines a residual-based update that moves the current iterate toward a solution of the estimating equation, an inertial extrapolation term that uses information from the preceding two iterations for acceleration, and a thresholding step that promotes sparse iterates. 
When $H$ is symmetric and positive semi-definite, thresholded inertial updates of this type can be viewed as accelerated proximal-gradient iterations for an $\ell^1$-regularized quadratic potential, and are closely related to the fast iterative shrinkage-thresholding algorithm of \citet{beck2009fast}.
\end{example}

For each fixed iteration index $t$, we regard $\beta_t$ in \eqref{eq:coef_pop} as an algorithm-induced population target in its own right. 
Indeed, once the initialization, tuning parameters, and update mappings $\{F_s\}_{s\leq t}$ are fixed, $\beta_t$ is a deterministic functional of the population quantities $(H,\xi)$, and hence of the underlying data-generating distribution. 
Although $\beta_t$ generally differs from the solution $\beta^\circ$ of $H\beta^\circ=\xi$, this discrepancy need not be viewed solely as an approximation error. 
Rather, the finite-iteration target incorporates the regularization imposed by the algorithm, including shrinkage, momentum, and early stopping, and may therefore represent a deliberately biased but more stable prediction rule. 
In high-dimensional or ill-conditioned problems, such algorithmic bias can reduce the sensitivity of the empirical iterate $\hat\beta_t$ to sampling fluctuations in weakly identified directions, and improve predictive performance relative to the unregularized solution \citep{raskutti2014early}. 
Accordingly, our inferential target is the iteration-specific parameter $\beta_t$, with $\hat\beta_t$ in \eqref{eq:coef_est} serving as its sample analogue, whereas convergence of $\beta_t$ as $t$ increases is a separate algorithmic consideration and is not required for $\beta_t$ to have a meaningful statistical interpretation.

\subsection{Algorithm-aware inference via bootstrap}
\label{sec:inference}
The common sampling perturbation enters the algorithm \eqref{eq:coef_est} at every iteration, so we first extract such perturbation along algorithmic propagation. 
To that end, assume $F_t$ to be absolutely continuous with Jacobian matrix $\nabla F_t = (F_{t,0},F_{t,1},\dots,F_{t,2t-1})$ in a block-row form, where $F_{t,k}:\R^{2tp}\to\R^{p\times p}$ spans columns $kp+1$ through
$(k+1)p$. 
The even and odd blocks multiply parameter and estimating-equation residual, respectively. 
\begin{example}[Jacobian matrix]\label{eg:Jacobian}
In the setting of Example~\ref{eg:alg}, $\nabla F_t(\theta_{t-1})\in\R^{p\times 2tp}$ is given as follows. Denote the pre-threshold values by 
\[z_t=(z_{t,1},\dots,z_{t,p})^\top=(1+\mu_t)(\beta_{t-1}-\eta_t\gamma_{t-1})-\mu_t(\beta_{t-2}-\eta_t\gamma_{t-2}).\]
Let $\Lambda_t=\operatorname{diag}(\1\{|z_{t,j}|>\lambda\eta_t\}:j=1,\dots,p)$ be the matrix indicating active coordinates. 
The only nonzero derivative blocks are 
\begin{align*}
F_{t,0}(\theta_{t-1})&=(1+\mu_t)\Lambda_t,
&F_{t,1}(\theta_{t-1})&=-(1+\mu_t)\eta_t\Lambda_t,\\
F_{t,2}(\theta_{t-1})&=-\mu_t\Lambda_t,
&F_{t,3}(\theta_{t-1})&=\mu_t\eta_t\Lambda_t.
\end{align*}
\end{example}

By the integral form of the mean value theorem, the difference between \eqref{eq:coef_est} and \eqref{eq:coef_pop} reads 
\begin{equation}\label{eq:linear}
\hat\beta_t-\beta_t=\tilde{D}_t(\hat\theta_{t-1}-\theta_{t-1})=D_t(\hat\theta_{t-1}-\theta_{t-1})+(\tilde{D}_t-D_t)(\hat\theta_{t-1}-\theta_{t-1}),
\end{equation}
where $\tilde{D}_t=\int_0^1\nabla F_t\big(\nu\hat\theta_{t-1}+(1-\nu)\theta_{t-1}\big)\dd{\nu}$ and $D_t=\nabla F_t(\theta_{t-1})$. 
The first term on the right-hand side of \eqref{eq:linear} is the leading linearized perturbation, while the second term is the corresponding nonlinear remainder. 
Under suitable smoothness and stability conditions, $\tilde{D}_t-D_t$ becomes negligible when $\hat\theta_{t-1}$ is close to $\theta_{t-1}$, making the remainder of smaller order than the leading term. 
Thus, to first order, statistical uncertainty in the empirical inputs is propagated along the algorithmic trajectory through the Jacobian matrices of the update mappings. 
It is notable that uncertainty generated at an earlier iteration affects subsequent iterates both directly, through the previous parameters, and indirectly, through the associated estimating-equation residuals. 

The linearization in \eqref{eq:linear} suggests an algorithm-aware bootstrap that perturbs the empirical inputs and recursively propagates the resulting perturbations through estimated Jacobian matrices of the algorithm. 
Let $\hat{D}_t=\nabla F_t(\hat\theta_{t-1})$ and let $(\hat{H}^*,\hat\xi^*)$ denote a bootstrap version of
$(\hat{H},\hat\xi)$. 
Starting from $\hat\beta_0^*=\hat\beta_0=\beta_0$, define the bootstrap trajectory by 
\begin{equation}\label{eq:coef_boot}
\hat\beta_t^*=\hat\beta_t+\hat{D}_t(\hat\theta_{t-1}^*-\hat\theta_{t-1}),\quad
\hat\theta_{t-1}^*=\big((\hat\beta_{t-1}^*)^\top,(\hat\gamma_{t-1}^*)^\top,\dots,(\hat\beta_0^*)^\top,(\hat\gamma_0^*)^\top\big)^\top,
\end{equation}
where $\hat\gamma_t^*=\hat{H}^*\hat\beta_t^*-\hat\xi^*$ for $t\in\bN\cup\{0\}$. 
In particular, $\hat\gamma_0^*$ differs from $\hat\gamma_0$, so the bootstrap perturbation enters at the initial residual even though the parameter initialization is fixed. 
The construction \eqref{eq:coef_boot} does not simply rerun the original nonlinear algorithm using the bootstrap inputs. Instead, it holds the empirical trajectory and its local Jacobian matrices fixed, and propagates bootstrap perturbations through the first-order dynamics of the algorithm. 
Therefore, \eqref{eq:coef_boot} preserves the temporal dependence induced by the iterative procedure while targeting the leading linearized fluctuation of $\hat\beta_t-\beta_t$. 
Under conditions ensuring that the Jacobian estimators are consistent and the accumulated nonlinear remainders are negligible, the conditional distribution of $\hat\beta_t^*-\hat\beta_t$ can consequently approximate the sampling distribution of $\hat\beta_t-\beta_t$. 

We next illustrate how the algorithm-aware bootstrap can be used to conduct simultaneous inference along the population trajectory \eqref{eq:coef_pop}.
Let $\mathcal{U}\subset\R^p$ be a collection of contrast directions and let $\mathcal{T}\subset\bN$ be a collection of iteration indices. 
Without loss of generality, we assume that $\mathcal{U}$ is symmetric, in the sense that $\mathcal{U}=\bigcup\{\pm u:u\in\mathcal{U}^+\}$ for a representative subcollection $\mathcal{U}^+$. 
Our goal is to construct simultaneous confidence bands for the iteration-specific contrasts \[\{u^\top\beta_t:(u,t)\in\mathcal{U}^+\times\mathcal{T}\}\] with joint confidence level $1-\alpha$.
When $\mathcal{T}$ is finite, the procedure is implemented as follows. 
First, compute the observed trajectory $(\hat\beta_t)_{t\in\mathcal{T}}$ according to \eqref{eq:coef_est}. 
Next, for each bootstrap replication $b=1,\dots,N$, independently generate a bootstrap version $(\hat{H}^{*b},\hat\xi^{*b})$ of $(\hat{H},\hat\xi)$, conditionally on the observed data, and compute the corresponding algorithm-aware bootstrap trajectory $(\hat\beta_t^{*b})_{t\in\mathcal{T}}$ using \eqref{eq:coef_boot}.
Given sample size $n$, define the maximal bootstrap statistic 
\[S_n^{*b}=\sqrt{n}\sup_{u\in\mathcal{U},t\in\mathcal{T}}u^\top(\hat\beta_t^{*b}-\hat\beta_t)=\sqrt{n}\sup_{u\in\mathcal{U}^+,t\in\mathcal{T}}\big|u^\top(\hat\beta_t^{*b}-\hat\beta_t)\big|,\]
which captures the largest absolute bootstrap fluctuation over all contrasts and iterations under consideration.
Let $c_{1-\alpha}^{*,N}$ denote the empirical $(1-\alpha)$ quantile of
$\{S_n^{*b}:1\le b\le N\}$. We then report the simultaneous confidence bands
\begin{equation}\label{eq:conf-band}
\big[u^\top\hat\beta_t\pm c_{1-\alpha}^{*,N}/\sqrt{n}\big],
\qquad(u,t)\in\mathcal{U}^+\times\mathcal{T}.
\end{equation}
The use of a common critical value accounts jointly for multiplicity across both the contrast directions and the iteration indices. 
If $\mathcal{T}$ is infinite, the same construction provides the theoretical target, whereas numerical implementation requires either a tractable evaluation of the supremum or a finite truncation of the trajectory.

Some choices of $\mathcal{U}^+$ lead to familiar inferential settings. 
For a single fixed contrast $u\in\R^p$, taking $\mathcal{U}^+=\{u\}$ yields simultaneous inference for $u^\top\beta_t$ over
$t\in\mathcal{T}$. Choosing $u=e_{j_0}$, where $e_{j_0}$ is the $j_0$th standard basis vector, gives inference for the trajectory of a single coordinate $\beta_{t,j_0}$. 
Of particular interest is the choice $\mathcal{U}^+=\{e_1,\dots,e_p\}$, for which 
\[\sup_{u\in\mathcal{U}^+}\big|u^\top(\hat\beta_t^{*b}-\hat\beta_t)\big|=\normvec{\hat\beta_t^{*b}-\hat\beta_t},\]
and the resulting bands cope with every coordinate of
$\beta_t$ at every iteration $t\in\mathcal{T}$. 
More generally, the set $\mathcal{U}^+$ may contain scientifically meaningful linear contrasts, allowing the same procedure to accommodate simultaneous inference for selected coordinates, groups of coefficients, or predictions at pre-specified covariate values.

\section{Theoretical Guarantees}
\label{sec:theory}

\subsection{First-order expansion of algorithmic trajectories}
\label{sec:expansion}
We begin with a first-order expansion of the empirical trajectory \eqref{eq:coef_est} that makes explicit its dependence on the sampling fluctuations in the estimating equation. 
For notational convenience, denote by $\tilde{D}_{t,k}$ (resp.\ $D_{t,k}$, $\hat{D}_{t,k}$) the submatrix of $\tilde{D}_t$ (resp.\ $D_t$, $\hat{D}_t$) consisting of columns $kp+1$ through $(k+1)p$. Define
\[\zeta=\mqty(\operatorname{vec}(\hat{H}-H)\\\hat\xi-\xi)\in\R^d,\quad
d=p^2+p.\]
The perturbation $\zeta$ in the algorithmic inputs drives the fluctuations of the entire empirical trajectory. 
Indeed, using $(\hat{H}-H)\beta_t=(\beta_t^\top\otimes I_p)\operatorname{vec}(\hat{H}-H)$, we obtain the decomposition 
\[\hat\gamma_t-\gamma_t
=\hat{H}(\hat\beta_t-\beta_t)+(\hat{H}-H)\beta_t-(\hat\xi-\xi)
=\hat{H}(\hat\beta_t-\beta_t)+\{(\beta_t^\top,-1)\otimes I_p\}\zeta.\]
Recall from \eqref{eq:linear} that $\hat\beta_t-\beta_t=\tilde{D}_t(\hat\theta_{t-1}-\theta_{t-1})$.
Partitioning the Jacobian matrix according to the parameter and residual components of the algorithmic history gives 
\begin{align*}
\hat\beta_t-\beta_t
&=\sum_{r=1}^t\big\{\tilde{D}_{t,2(r-1)}(\hat\beta_{t-r}-\beta_{t-r})+\tilde{D}_{t,2r-1}(\hat\gamma_{t-r}-\gamma_{t-r})\big\}\\
&=\sum_{r=1}^t\{\tilde{D}_{t,2(r-1)}+\tilde{D}_{t,2r-1}\hat{H}\}(\hat\beta_{t-r}-\beta_{t-r})+\sum_{r=1}^t\tilde{D}_{t,2r-1}\{(\beta_{t-r}^\top,-1)\otimes I_p\}\zeta.
\end{align*}
To identify the leading linear term, define the sensitivity matrices $\Pi_t\in\R^{p\times d}$ recursively by $\Pi_0=0_{p\times d}$ and
\begin{equation}\label{eq:sensitivity}
\Pi_t=\sum_{r=1}^t\{D_{t,2(r-1)}+D_{t,2r-1}H\}\Pi_{t-r}+\sum_{r=1}^tD_{t,2r-1}\{(\beta_{t-r}^\top,-1)\otimes I_p\}.
\end{equation}
We think of $\Pi_t$ as the derivative of the $t$th population iterate \eqref{eq:coef_pop} with respect to a local perturbation of $(H,\xi)$, propagated through the preceding algorithmic updates. 
Therefore, $\Pi_t\zeta$ is the candidate first-order approximation to $\hat\beta_t-\beta_t$. 
The recursion \eqref{eq:sensitivity} also makes clear that $\Pi_t$ is algorithm-dependent. Even for the same population estimating equation, different initializations, tuning parameters, or update mappings may produce different sensitivities and, consequently, different finite-iteration sampling distributions.

We now introduce some quantities that control the accumulation of perturbations and nonlinear approximation errors. 
Let $\bar{S}\subset\{1,\dots,p\}$ be a reference set of size $s$, and let \[Q=\operatorname{diag}(\1\{j\in\bar{S}\}:j=1,\dots,p)\] be the associated coordinate projection. 
Denote 
\[a_t=\sum_{r=1}^t\normop{\{D_{t,2(r-1)}+D_{t,2r-1}H\}Q},\qand
b_t=\sum_{r=1}^t\normop{D_{t,2r-1}}.\]
Here, $a_t$ measures the local amplification of perturbations in preceding parameter iterates, restricted to the reference coordinates, while $b_t$ measures the aggregate sensitivity of the $t$th update to preceding residuals. 
The quantity \[Z=\normop{H Q}+s\normvec{\zeta}\] controls the $\ell^\infty$ operator norm of $\hat{H}$ restricted to vectors supported on $\bar{S}$. 
For $\epsilon>0$, define the local Jacobian-continuity moduli 
\begin{align*}
\omega_{t,0}(\epsilon)&=\sup\bigg\{\sum_{r=0}^{t-1}\normop{\{F_{t,2r}(\tilde\theta_{t-1})-D_{t,2r}\}Q}:\tilde\theta_{t-1}\in\R^{2tp},\normvec{\tilde\theta_{t-1}-\theta_{t-1}}\le\epsilon\bigg\},\\
\omega_{t,1}(\epsilon)&=\sup\bigg\{\sum_{r=1}^t\normop{F_{t,2r-1}(\tilde\theta_{t-1})-D_{t,2r-1}}:\tilde\theta_{t-1}\in\R^{2tp},\normvec{\tilde\theta_{t-1}-\theta_{t-1}}\le\epsilon\bigg\}.
\end{align*}
These quantities measure the maximal variation of the parameter and residual blocks of the Jacobian over an $\epsilon$-neighborhood of the population history. 
For differentiable mappings $F_t$ with locally continuous Jacobians, $\omega_{t,0}(\epsilon)$ and $\omega_{t,1}(\epsilon)$ tend to zero as $\epsilon\to0$. They vanish identically when $F_t$ is affine on the relevant neighborhood, which is the case in Example~\ref{eg:Jacobian}.

The following Theorem~\ref{thm:linear} provides a finite-sample bound for the remainder in the first-order expansion of the empirical trajectory. 
\begin{theorem}\label{thm:linear}
Let $T\in\bN$ and $\epsilon>0$.
Define $\delta_0=0$ and 
\begin{align*}
\delta_t&=\max[\delta_{t-1},
\{a_t+b_ts\normvec{\zeta}+(\omega_{t,0}+Z\omega_{t,1})(\epsilon)\}\delta_{t-1}\\
&\hspace{5em}+\normvec{\Pi_t\zeta}+a_t\max_{r\le t-1}\normvec{\Pi_r\zeta}+\omega_{t,1}(\epsilon)(\max_{r\le t-1}\normsum{\beta_r}+1)\normvec{\zeta}],
\end{align*}
recursively for $t\in\bN$. 
Suppose that the following event holds: for all $t=0,1,\dots,T-1$, 
\begin{enumerate}[label=(\roman*)]
\setlength{\itemsep}{1ex}
\item\label{cond:small} $\max(Z,1)\delta_t+(\normsum{\beta_t}+1)\normvec{\zeta}\le\epsilon$, and 
\item\label{cond:support} $\operatorname{supp}(\beta_t)\cup\operatorname{supp}(\hat\beta_t)\cup\operatorname{supp}(\Pi_t\zeta)
\subset\bar{S}$.
\end{enumerate}
Then for all $t=1,\dots,T$, 
\begin{align*}
\normvec{\hat\beta_t-\beta_t-\Pi_t\zeta}
&\le a_t\max_{r\le t-1}\normvec{\hat\beta_r-\beta_r-\Pi_r\zeta}
+\omega_{t,1}(\epsilon)(\max_{r\le t-1}\normsum{\beta_r}+1)\normvec{\zeta}\\
&\quad+\{b_ts\normvec{\zeta}+(\omega_{t,0}+Z\omega_{t,1})(\epsilon)\big\}\delta_{t-1}.
\end{align*}
\end{theorem}

The sequence $(\delta_t)$ is constructed to control the perturbation of the algorithmic history up to iteration $t$. More precisely, its recursive definition accounts for both the leading linear fluctuations and the accumulation of nonlinear errors through successive updates. 
When $\normvec{\zeta}$ is of first order and the local Jacobian-continuity moduli $\omega_{t,j}(\epsilon)$ are of a comparable or smaller order on the relevant neighborhoods, $\delta_t$ is likewise of first order, provided that the algorithmic amplification factors remain controlled. 
Condition~\ref{cond:small} then closes the local argument by ensuring that the empirical history remains within the $\epsilon$-neighborhood of the population history. 
Condition~\ref{cond:support} requires the population iterates, the empirical iterates, and their leading linear approximations to remain supported on the reference set. This support restriction permits the perturbation analysis to depend on the effective dimension $s$ rather than directly on the ambient dimension $p$.

The final bound in Theorem~\ref{thm:linear} shows the linearization that 
\[\hat\beta_t-\beta_t=\Pi_t\zeta+\text{higher-order remainder}.\]
Here, $a_t$ serves as a local algorithmic stability coefficient that controls the extent to which linearization errors accumulated at earlier iterations are transmitted to the $t$th update. 
The remaining terms arise from the interaction between the sampling fluctuations and the preceding trajectory errors, as well as from local variation of the Jacobian of $F_t$. 
Thus, the linearization error is of smaller order than the leading term $\Pi_t\zeta$, when $\normvec{\zeta}$ is small, the moduli $\omega_{t,j}(\epsilon)$ are negligible on shrinking neighborhoods, and the coefficients are well bounded. 
This conclusion may hold not only at each fixed iteration, but also uniformly over a collection of iterations allowed to grow with the sample size. 
A particularly strong form of stability is obtained if 
\[\limsup_{t\to\infty}a_t<1,\quad
\sup_{t\in\bN}b_t<\infty,\quad
\sup_{t\in\bN}\normsum{\beta_t}<\infty,\]
\[\qand\lim_{\epsilon\to0}\limsup_{t\to\infty}\omega_{t,j}(\epsilon)=0,\quad j\in\{0,1\}.\]
Under these conditions, and provided that the local-neighborhood and support events hold uniformly, the linearization error can be controlled uniformly over $t\in\bN$. 
For the update considered in Examples~\ref{eg:alg}--\ref{eg:Jacobian}, the coefficients take the form 
\[a_t=(1+2\mu_t)\normop{\Lambda_t(I_p-\eta_tH)Q}\le(1+2\mu_t)\normop{Q(I_p-\eta_tH)Q},\]
\[b_t=(1+2\mu_t)\eta_t\normop{\Lambda_t}\le(1+2\mu_t)\eta_t,\]
and the moduli $\omega_{t,j}(\epsilon)$ can be bounded in terms of the discrepancy between the active set at the population history and the active sets attained within its $\epsilon$-neighborhood. 

In what follows, we establish a bootstrap counterpart of Theorem~\ref{thm:linear}.
Conditionally on the observed data, define 
\[\zeta^*=\mqty(\operatorname{vec}(\hat{H}^*-\hat{H})\\\hat\xi^*-\hat\xi),\]
which represents the additional perturbation introduced by the bootstrap.
In particular, the total displacement of the bootstrap inputs from their population counterparts is governed by both $\zeta$ and $\zeta^*$. 
To provide the corresponding bound for $\normop{\hat{H}^*Q}$, let \[Z^*=Z+s\normvec{\zeta^*}=\normop{H Q}+s(\normvec{\zeta}+\normvec{\zeta^*}).\]
The algorithm-aware bootstrap \eqref{eq:coef_boot} is constructed to reproduce the first-order response of the empirical trajectory to the additional perturbation. 
The following Theorem~\ref{thm:linear_boot} shows that its leading term is $\Pi_t\zeta^*$, also involving the population sensitivity matrix $\Pi_t$ in \eqref{eq:sensitivity}.
\begin{theorem}\label{thm:linear_boot}
Work under the setup and conditions of Theorem~\ref{thm:linear}. 
Define $\delta_0^*=0$ and 
\begin{align*}
\delta_t^*&=\max\big(\delta_{t-1}^*,
\{a_t+b_ts(\normvec{\zeta}+\normvec{\zeta^*})+(\omega_{t,0}+Z^*\omega_{t,1})(\epsilon)\}\delta_{t-1}^*+\normvec{\Pi_t\zeta^*}\\
&\hspace{3em}+a_t\max_{r\le t-1}\normvec{\Pi_r\zeta^*}+[\omega_{t,1}(\epsilon)(\max_{r\le t-1}\normsum{\beta_r}+1)+\{b_t+\omega_{t,1}(\epsilon)\}s\delta_{t-1}]\normvec{\zeta^*}\big),
\end{align*}
recursively for $t\in\bN$. 
Suppose further that for all $t=0,1,\dots,T-1$, 
\[\operatorname{supp}(\hat\beta_t^*)\cup\operatorname{supp}(\Pi_t\zeta^*)\subset\bar{S}.\]
Then for all $t=1,\dots,T$, 
\begin{align*}
\normvec{\hat\beta_t^*-\hat\beta_t-\Pi_t\zeta^*}
&\le a_t\max_{r\le t-1}\normvec{\hat\beta_r^*-\hat\beta_r-\Pi_r\zeta^*}\\
&\quad+\{b_ts(\normvec{\zeta}+\normvec{\zeta^*})+(\omega_{t,0}+Z^*\omega_{t,1})(\epsilon)\}\delta_{t-1}^*\\
&\quad+[\omega_{t,1}(\epsilon)(\max_{r\le t-1}\normsum{\beta_r}+1)+\{b_t+\omega_{t,1}(\epsilon)\}s\delta_{t-1}]\normvec{\zeta^*}.
\end{align*}
\end{theorem}

The sequence $(\delta_t^*)$ plays the same role for the bootstrap trajectory as $(\delta_t)$ does for the empirical trajectory. 
It controls the deviation of the bootstrapped parameter from the estimator up to iteration $t$, given that the empirical trajectory has itself been perturbed away from the population trajectory. 
This explains the simultaneous appearance of $\zeta$ and $\zeta^*$. 
The final bound in Theorem~\ref{thm:linear_boot} has a structure parallel to that of Theorem~\ref{thm:linear}. 
The first term propagates bootstrap approximation errors from earlier iterations, with $a_t$ again acting as the local algorithmic stability coefficient. 
The second term controls interactions between the accumulated bootstrap trajectory error and the empirical and bootstrap perturbations of the estimating equation. 
The third term reflects the fact that the bootstrap expansion is centered at the empirical trajectory. 
If $\zeta$ and $\zeta^*$ conditional on the data are both of first order, the local Jacobian-continuity moduli are small, and the algorithmic stability coefficients remain controlled, then Theorem~\ref{thm:linear_boot} yields 
\[\hat\beta_t^*-\hat\beta_t=\Pi_t\zeta^*+\text{higher-order bootstrap remainder}.\]
Together with the original expansion, this implies that the sampling and bootstrap fluctuations are governed by the same linear operator $\Pi_t$. 
Such a common influence structure is the central mechanism through which the algorithm-aware bootstrap reproduces the joint distribution of the empirical trajectory.

\subsection{Gaussian approximation and bootstrap validity}
\label{sec:Gauss}
We now combine the trajectory expansions with Gaussian and multiplier couplings for suprema of empirical processes.
The deterministic index sets $\mathcal{U}=\mathcal{U}_n$ and $\mathcal{T}=\mathcal{T}_n$ may depend on $n$ and need not be finite.

Let $W,W_1,\dots,W_n$ be independent and identically distributed random elements with common distribution $P_W$. 
Suppose that $\zeta$ admits the asymptotically linear representation 
\[\zeta=n^{-1}\sum_{i=1}^n\ell(W_i)+R_n,\qquad
\E\{\ell(W_i)\}=0_d,\qquad \Sigma=\E\{\ell(W_i)\ell(W_i)^\top\}.\]
For a random vector $g\sim\N(0_d,I_d)$, define the Gaussian process 
\[G(u,t)=u^\top\Pi_t\Sigma^{1/2}g,\quad(u,t)\in\mathcal{U}\times\mathcal{T}.\]
Consider the maximal empirical and Gaussian statistics 
\[S_n=\sqrt{n}\sup_{u\in\mathcal{U},t\in\mathcal{T}}u^\top(\hat\beta_t-\beta_t),\qquad
S^G=\sup_{u\in\mathcal{U},t\in\mathcal{T}}G(u,t).\]
Since $\mathcal{U}$ is symmetric, these suprema are also suprema of absolute contrasts.
The following Lemma~\ref{lem:Gauss} gives sufficient conditions for the distributional approximation of $S_n$ to $S^G$. 
Define the effective class of influence scores by
\[\mathcal{F}_n=\{f_{u,t}=u^\top\Pi_t\ell:(u,t)\in\mathcal{U}\times\mathcal{T},u^\top\Pi_t\ne0_d^\top\}.\]
Omitting structural zero directions from $\mathcal{F}_n$ leaves both linear suprema unchanged.
Let $\bar f_n$ be a measurable envelope for $\mathcal{F}_n$, and write $P_W h=\E\{h(W)\}$.

\begin{lemma}\label{lem:Gauss}
Suppose that for constants $A_n\ge \mathrm{e}$ and $v_n\ge1$, 
\begin{equation}\label{eq:cck-entropy}
\sup_Q N\{\varepsilon(Q\bar f_n^2)^{1/2},\mathcal{F}_n,L^2(Q)\}\le(A_n/\varepsilon)^{v_n},\qquad 0<\varepsilon\le1,
\end{equation}
where the supremum is over all finitely supported probability measures $Q$.
Assume that, for fixed $0<\underline\sigma\le\overline\sigma<\infty$, a fixed $q\in[4,\infty)$ and a deterministic $B_n\ge\max(\overline\sigma,1)$,
\begin{equation}\label{eq:cck-moments}
\inf_{f\in\mathcal{F}_n}P_W f^2\ge\underline\sigma^2,\qquad
\sup_{f\in\mathcal{F}_n}P_W|f|^k\le\overline\sigma^2B_n^{k-2},\quad k=2,3,4,\qquad
P_W|\bar f_n|^q\le B_n^q.
\end{equation}
Set $K_n=v_n\log\max(A_nB_n/\overline\sigma,n)$. 
Suppose that
\begin{equation}\label{eq:moment-cond}
\frac{B_nK_n^{3/2}}{n^{1/2-1/q}}+\frac{B_n^{1/3}K_n^{7/6}}{n^{1/6}}\to0
\end{equation}
and
\begin{equation}\label{eq:linear-cond}
\sup_{(u,t)\in\mathcal{U}\times\mathcal{T}}\abs{u^\top(\hat\beta_t-\beta_t-\Pi_t\zeta)}
+\sup_{(u,t)\in\mathcal{U}\times\mathcal{T}}\abs{u^\top\Pi_tR_n}
=o_{\Pr}\{(nK_n)^{-1/2}\}.
\end{equation}
Then
\[\sup_{x\in\R}\abs{\Pr(S_n\le x)-\Pr(S^G\le x)}\to0.\]
\end{lemma}

The proof follows from \citet[Theorem~2.1]{chernozhukov2016couplings} and the anti-concentration bound in \citet[Lemma~A.1]{chernozhukov2014suprema}.
Condition~\eqref{eq:cck-entropy} controls the complexity of the entire class, while condition~\eqref{eq:linear-cond} requires the nonlinear trajectory remainder and the influence-function remainder to be negligible relative to the scale of the Gaussian statistic. 
The Supplementary Material derives covering bounds and uniform remainder rates for verifying the conditions.

We next introduce the multiplier bootstrap. 
Let $\varpi_1,\dots,\varpi_n$ be independent standard normal random variables that are independent of the observed data. 
Let $\hat\ell_1,\dots,\hat\ell_n$ be estimated influence vectors satisfying $\sum_{i=1}^n\hat\ell_i=0_d$, and set $(\hat{H}^*,\hat\xi^*)$ so that 
\[\zeta^*=n^{-1}\sum_{i=1}^n\varpi_i\hat\ell_i.\]
Lemma~\ref{lem:bootstrap} below shows that the conditional distribution of 
\[S_n^*=\sqrt{n}\sup_{u\in\mathcal{U},t\in\mathcal{T}}u^\top(\hat\beta_t^*-\hat\beta_t).\]
is approximated by the same Gaussian supremum law as the sampling distribution of $S_n$.
Write $\bar\ell=n^{-1}\sum_i\ell(W_i)$ and define
\[J_n=\sup_{(u,t)\in\mathcal{U}\times\mathcal{T}}\normsum{u^\top\Pi_t},\qquad
e_n=\max_{1\le j\le d}\Big[n^{-1}\sum_{i=1}^n\{\hat\ell_{ij}-\ell_j(W_i)+\bar\ell_j\}^2\Big]^{1/2}.\]

\begin{lemma}\label{lem:bootstrap}
Suppose that the conditions of Lemma~\ref{lem:Gauss} hold and that
\begin{equation}\label{eq:cck-score-rate}
J_ne_nK_n^{1/2}\log^{1/2}(2d)=o_{\Pr}(1).
\end{equation}
Assume also that
\begin{equation}\label{eq:linear_boot-cond}
\sup_{(u,t)\in\mathcal{U}\times\mathcal{T}}\abs{u^\top(\hat\beta_t^*-\hat\beta_t-\Pi_t\zeta^*)}
=o_{\Pr_{|W_1^n}}\{(nK_n)^{-1/2}\}
\end{equation}
in probability.
Then
\[\sup_{x\in\R}\abs{\Pr_{|W_1^n}(S_n^*\le x)-\Pr(S^G\le x)}=o_{\Pr}(1).\]
\end{lemma}

The proof first applies \citet[Theorem~2.2]{chernozhukov2016couplings} to the multiplier process with the centered true influence functions, and then controls the error introduced by estimating those functions.
Condition~\eqref{eq:cck-score-rate} is an explicit sufficient condition for this second step. For exact sample-average inputs and their empirically centered contributions, $R_n=0_d$ and $e_n=0$.
Condition~\eqref{eq:linear_boot-cond} is the bootstrap counterpart of \eqref{eq:linear-cond}. 

Combining the two Gaussian approximations yields bootstrap validity.
\begin{theorem}\label{thm:valid}
Under the conditions of Lemma~\ref{lem:bootstrap},
\[\sup_{x\in\R}\abs{\Pr_{|W_1^n}(S_n^*\le x)-\Pr(S_n\le x)}=o_{\Pr}(1).\]
For any fixed $\alpha\in(0,1)$, let
\[c_{1-\alpha}^*=\inf\{x:\Pr_{|W_1^n}(S_n^*\le x)\ge1-\alpha\}.\]
Then \[\Pr(S_n\le c_{1-\alpha}^*)\to1-\alpha.\]
\end{theorem}
Theorem~\ref{thm:valid} guarantees that the confidence bands \eqref{eq:conf-band} based on $c_{1-\alpha}^*$ attain asymptotically correct simultaneous coverage over $\mathcal{U}\times\mathcal{T}$.

\section{Numerical Studies}
\label{sec:numeric}

\subsection{Simulation}
\label{sec:simu}
We evaluate the proposed method in three estimating equations in Example~\ref{eg:LEE}. 
In every experiment, fix $p=1000$ and $\beta^\circ=(1,-0.8,0.6,0,\dots,0)^\top$, and let $n\in\{300,600,1000\}$.
Write $V_\rho\in\R^{p\times p}$ for the matrix having entries $(V_\rho)_{jk}=\rho^{|j-k|}$, where $\rho\in\{0,0.05,0.10,0.40,0.70\}$.
The data are sampled as follows:
\begin{enumerate}
\item\emph{Linear regression.}
$x_i\sim\N(0_p,V_\rho)$ and
$y_i=x_i^\top\beta^\circ+\varepsilon_i$;
\item\emph{Paired Fisher discrimination.}
$x_i^\pm=\pm0.5V_\rho\beta^\circ+\sqrt{0.5}V_\rho^{1/2}z_i^0+\sqrt{0.5}V_\rho^{1/2}z_i^\pm$;
\item\emph{Least-squares temporal-difference learning (with discount factor $\delta=0.7$).}
$x_i\sim\N(0_p,V_\rho)$, $x_i'=A^\top x_i+0.4V_\rho^{1/2}z_i$ where $A\in\R^{p\times p}$ has diagonal entries $0.25$ and first super-diagonal entries $0.10$, and $y_i=(x_i-\delta x_i')^\top\beta^\circ+\varepsilon_i$.
\end{enumerate}
The regression error $\varepsilon_i$, the pair-specific Fisher innovations $z_i^0,z_i^+,z_i^-$, and the temporal-difference transition and reward innovations $(z_i,\varepsilon_i)$ are distributed as standard normal variables or $t_5$ variables standardized to have unit variance.
This changes the tail behavior while preserving the estimating equation. 
We run the inertial thresholding algorithm in Example~\ref{eg:alg} with $\lambda=0.15\sqrt{1000/n}$, and cap the inertial coefficient $\mu_t$ at $0.10$. The step sizes corresponding to $\rho=0,0.05,0.10,0.40,0.70$ are $\eta_t=0.60,0.60,0.60,0.35,0.15$, respectively. 
Each setup defines its own population algorithmic target path. 
We consider the first-coordinate class $\mathcal{U}=\{\pm e_1\}$ and the all-coordinate class $\mathcal{U}=\{\pm e_j:1\le j\le p\}$.
The reported time set is $\mathcal{T}=\{5,\dots,40,80\}$, with $80$ serving as the numerical terminal iterate. The deviation from the full discrete grid $\{5,\dots,80\}$ is small enough to neglect.
For each configuration, we use $2000$ Monte Carlo samples and $999$ Gaussian multiplier draws per sample. 

\begin{figure}[!ht]
\centering
\includegraphics[width=\linewidth]{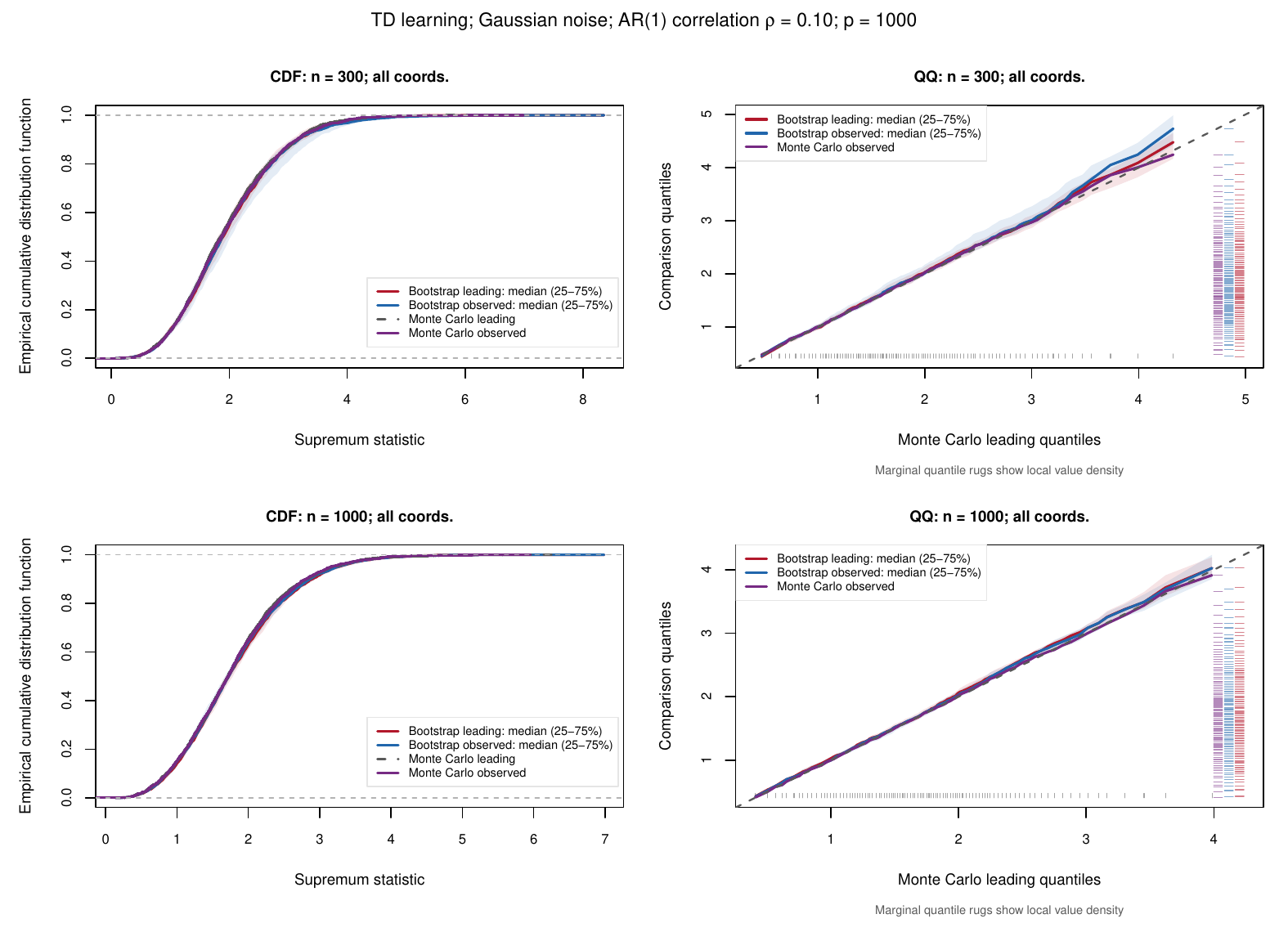}
\caption{Distributional diagnostics for all-coordinate inference in temporal-difference learning with Gaussian noise and $\rho=0.10$. 
The sample size is $n=300$ (top) and $n=1000$ (bottom). 
``Observed'' denotes the nonlinear trajectory statistic and ``leading'' denotes the first-order statistic, whose differences display linearization error. 
The bootstrap leading and observed curves show the median and middle 50\% across $20$ anchor data sets.
The differences between the Monte Carlo and bootstrap leading laws display the Gaussian approximation used for calibration. 
Marginal quantile rugs show local value density.}
\label{fig:sim-dist-main}
\end{figure}

We first examine the approximation mechanism before reporting coverage.
Figure~\ref{fig:sim-dist-main} considers all-coordinate inference for the non-symmetric temporal-difference learning under Gaussian noise and $\rho=0.10$.  
It compares the Monte Carlo distributions of the observed and leading statistics with the corresponding anchor-specific bootstrap distributions.  
The discrepancy between the observed and leading statistics measures the effect of algorithmic linearization, whereas the discrepancy between the Monte Carlo leading statistics and the Gaussian-multiplier bootstrap leading statistics assesses the Gaussian approximation used by the bootstrap. 
Comparing the two observed laws then displays the combined approximation used for inference.
At $n=300$, the observed Monte Carlo distribution differs visibly from its
leading approximation in the upper tail, possibly due to the scarce extremes, while the bootstrap leading law is already close to the Monte Carlo leading law.  
At $n=1000$, the observed and leading curves are close over most of their range.  Thus the main finite-sample discrepancy in this example comes from the nonlinear trajectory rather than from Gaussian approximation of the leading term.

\begin{figure}[!ht]
\centering
\includegraphics[width=\linewidth]{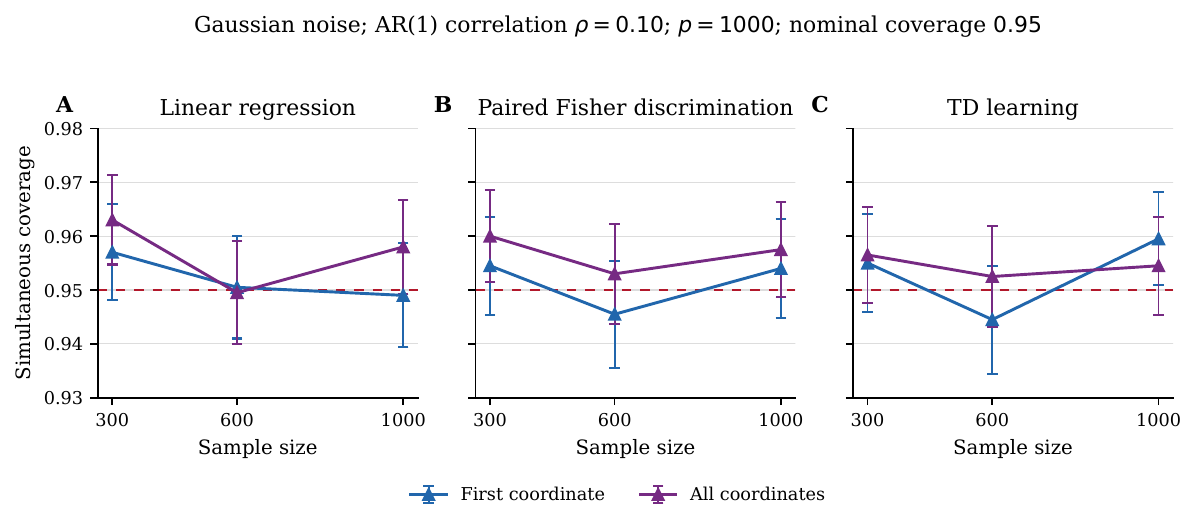}
\caption{Simultaneous coverage in a representative stable configuration.
The error bars are pointwise $95\%$ Monte Carlo intervals based on $2000$
replications. 
Each point is calibrated jointly over $\mathcal{T}=\{5,\dots,40,80\}$ and over either the first coordinate or all $p=1000$ coordinates.}
\label{fig:sim-coverage-main}
\end{figure}

Figure~\ref{fig:sim-coverage-main} shows simultaneous coverage in the same
stable Gaussian setting. 
Across the three estimating equations, the displayed estimates range from $0.9445$ to $0.9630$. 
Their Monte Carlo intervals generally overlap the nominal level $0.95$, although the all-coordinate procedure is mildly conservative in several configurations.
The complete coverage grid is reported in the Supplementary Material.  
Among all Gaussian and Student-$t_5$ configurations with $\rho\le0.10$, for which the implemented population diagnostic satisfies $\sup_ta_t<1$, the first- and all-coordinate coverage estimates range from $0.9415$ to $0.9705$ and have mean $0.9551$. 
At $\rho=0.40$ and $0.70$, coverage remains close to nominal in many runs, but $\sup_ta_t>1$ and several all-coordinate linearization and truncation errors are large. 
These stress tests show that coverage can be empirically adequate outside the sufficient stability condition, although they do not validate the time-uniform regime. 
Note that the mean coordinate-iteration active-set agreement exceeds $0.9997$ in such stable configurations, partly verifying the support events in Theorems~\ref{thm:linear}--\ref{thm:linear_boot}.

\subsection{Real data example}
\label{sec:real-data}
We apply the proposed simultaneous inference procedure to the Drink Less micro-randomized trial \citep{bell2023notifications}. 
The public data and study code are available from \url{https://osf.io/w3szp/}. 
The data contain $n=349$ participants observed at $30$ daily randomization times. 
At each time, the app assigned no notification with probability $0.4$, a standard notification with probability $0.3$, or a new notification with probability $0.3$. 
The proximal outcome $y_{ij}$ is the indicator that records whether the participant~$i$ opened the app during the subsequent one-hour window on day~$j$. 

Write $A_{ij}\in\{0,1,2\}$ for the assignment of participant $i$ on day $j$,
where $0$, $1$, and $2$ denote no, standard, and new notification, respectively. 
We use the two probability-orthonormal treatment contrasts $c=(c_1,c_2)^\top$ defined as 
\[c(0)=\left(0,-\sqrt{3/2}\right)^\top,\qquad
c(1)=\left(\sqrt{5/3},\sqrt{2/3}\right)^\top,\qquad
c(2)=\left(-\sqrt{5/3},\sqrt{2/3}\right)^\top.\]
The coordinates compare standard with new notification and either notification with no notification. 
Under the randomization probabilities $(0.4,0.3,0.3)$, they satisfy
that $\E\{c(A_{ij})\}=0_2$ and $\E\{c(A_{ij})c(A_{ij})^\top\}=I_2$.
For $m$ ordered time points, define the type-II discrete-cosine basis
\[\phi_0^{(m)}(r)=1,\qquad
\phi_k^{(m)}(r)=\sqrt{2}\cos\!\left\{\frac{k\pi(r-1/2)}{m}\right\},\quad k=1,\ldots,m-1,\]
so that $m^{-1}\sum_{r=1}^m\phi_k^{(m)}(r)\phi_l^{(m)}(r)=\1\{k=l\}$. 
The feature vector $x_{ij}$ contains the 60 main effects
\[\phi_k^{(30)}(j)c_a(A_{ij}),\qquad k=0,\dots,29,\quad a=1,2,\]
and, for each lag $h=1,2,3$, the $4(30-h)$ interactions
\[\1\{j>h\}\sqrt{\frac{30}{30-h}}\,\phi_k^{(30-h)}(j-h)c_a(A_{ij})c_b(A_{i,j-h}),\quad k=0,\dots,29-h,\quad a,b=1,2.\]
Thus $p=60+4(29+28+27)=396$. 
The features are pre-specified, and the normalization of each lag-specific basis is chosen relative to all 30 decision times.  

We form the participant-level contributions
\[H_i=\frac{1}{30}\sum_{j=1}^{30}x_{ij}x_{ij}^\top,\qquad
\xi_i=\frac{1}{30}\sum_{j=1}^{30}y_{ij}x_{ij},\]
whose sample averages give $\hat{H}$ and $\hat\xi$.
This construction gives the population target a direct design-based interpretation. Because treatment assignments are randomized and the feature dictionary is orthonormal under the randomization distribution, $H=I_p$ and hence $\beta^\circ=\xi$. The main-effect coordinates of $\beta^\circ$ are discrete-cosine coefficients of the time-varying marginal proximal response contrasts. 
For a fixed initialization and fixed tuning parameters, applying the population recursion to $(H,\xi)$ defines $\beta_t$ as a deterministic functional of the trial distribution. Its coordinates may be interpreted as algorithmically regularized versions of the corresponding proximal response contrasts.
Running the inertial thresholding algorithm in Example~\ref{eg:alg}, we set $\lambda=0.020$, use step size $\eta=0.20$, cap the inertial coefficient at $0.10$, and report $t=5,\dots,40$ together with numerical terminal iterate $200$. 
The inferential statement treats these tuning choices as fixed in advance.
A bootstrap multiplier is generated once for each participant and applied to all 30 observations from that participant, preserving arbitrary within-participant dependence.

\begin{figure}[!ht]
\centering
\includegraphics[width=0.8\linewidth]{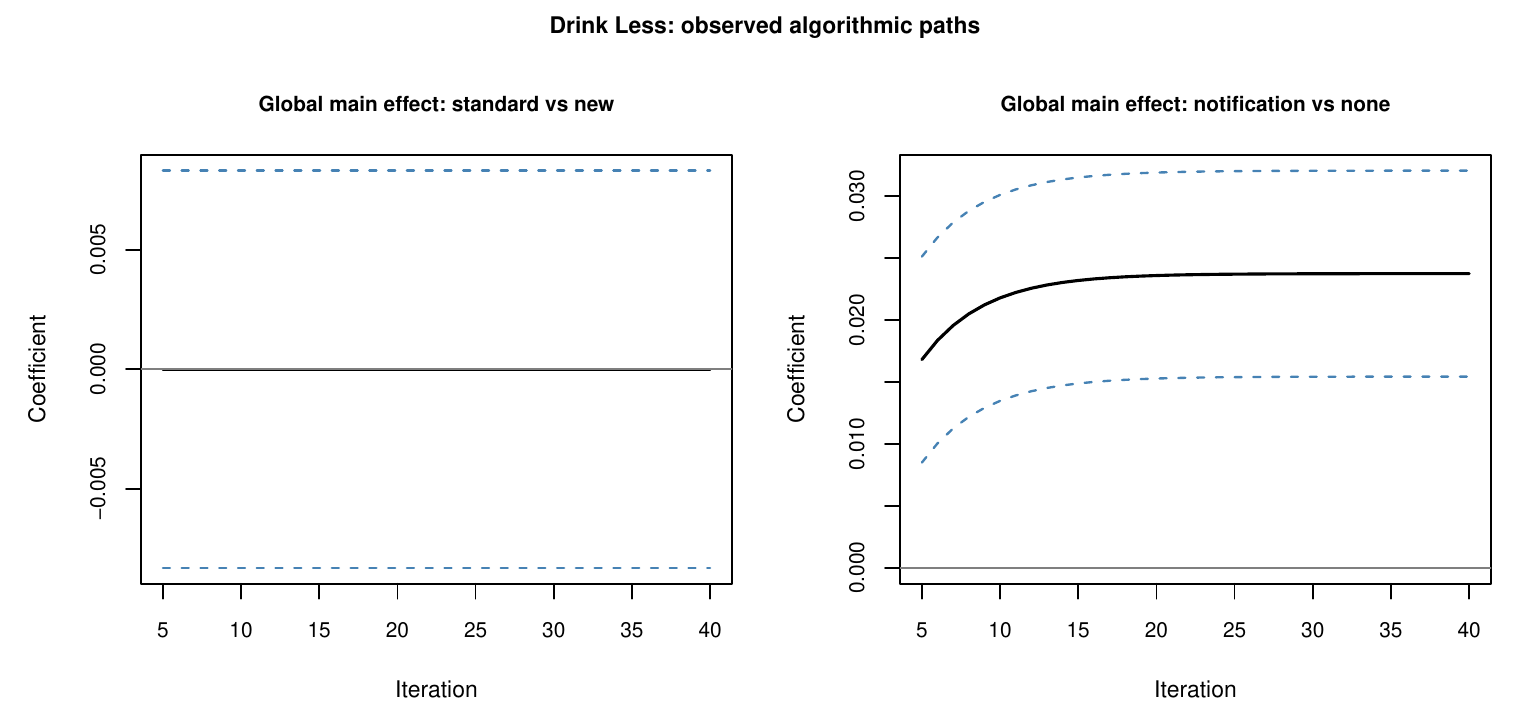}
\caption{Observed coefficient trajectories and simultaneous bands for the two global treatment contrasts in the Drink Less trial. 
The notification-versus-none band excludes zero throughout the reported iteration window, whereas the standard-versus-new path is thresholded to zero.}
\label{fig:drinkless-paths}
\end{figure}

The algorithm selects the constant notification-versus-none contrast as the only nonzero coordinate. Its endpoint estimate is $0.02374$, with the $95\%$ simultaneous interval $[0.01543,\ 0.03205]$. The band excludes zero at every reported iteration from $t=5$ onward, as seen in Figure~\ref{fig:drinkless-paths}. 
Because the treatment contrasts use probability-orthonormal coding, the coefficient $0.02374$ is not itself on the raw opening-probability scale. Multiplication by $\sqrt{2/3}-(-\sqrt{3/2})=5/\sqrt{6}$ gives the regularized notification-versus-none contrast $0.0485$, with simultaneous interval $[0.0315,\ 0.0654]$. These quantities target the algorithmically regularized counterpart of the day-averaged marginal response contrast and should not be equated with the raw, unregularized arm-mean difference. 
The constant standard-versus-new contrast, all non-constant main effects, and all lag interactions are thresholded to zero. Their inactivity describes the sparsity imposed by the algorithm and should not be interpreted as evidence that the corresponding unregularized population contrasts are exactly zero.

We provide additional evidence of the algorithm's stability.
Under $H=I_p$, the population lag-propagation diagnostic is $a_t\le0.96<1$, and the residual sensitivity diagnostic is $\sup_tb_t\le0.24$. 
The bootstrap critical value is $0.1552$, while the maximum discrepancy between the bootstrap supremum over $\{5,\dots,200\}$ and over the reported grid $\{5,\dots,40,200\}$ is $3.63\times10^{-8}$. 
As a descriptive sensitivity check, varying the fixed penalty over $\lambda\in\{0.010,0.015,0.020,0.025,0.030\}$ leaves the standard-versus-new coefficient at zero and the notification-versus-none coefficient positive, ranging from $0.01376$ to $0.03374$. 
This grid is not used to select $\lambda$ and carries no additional simultaneous inferential claim.


\section*{Acknowledgement}
Alexander Giessing is supported by the Ministry of Education, Singapore, under the Academic Research Fund Tier 1 (A-8004834-00-00).

\appendix

\section*{Supplementary material}
\label{SM}
The Supplementary Material contains proofs of all theoretical results and additional numerical results. 
The R code for producing numerical results is available in the GitHub repository \url{https://github.com/kellty/AlgLEE}.

\clearpage
\forcsvlist{\renewSformat}{assumption,lemma,theorem,corollary,equation,section,table,figure,algocf}

\begin{center}
{\LARGE Supplement to ``Algorithm-aware Statistical Inference in High-dimensional Estimating Equations''}\\[2ex]
{\large Tongyu Li and Alexander Giessing}\\[5ex]
\end{center}

The Supplementary Material contains proofs of all theoretical results and additional numerical results.

\section{Proofs of Theorems}

\subsection*{Proof of Theorem~\ref{thm:linear}}
By definition, for all $t\in\bN$,
\begin{align*}
\hat\beta_t-\beta_t-\Pi_t\zeta
&=\sum_{r=1}^t\{D_{t,2(r-1)}+D_{t,2r-1}H\}(\hat\beta_{t-r}-\beta_{t-r}-\Pi_{t-r}\zeta)\\
&\quad+\sum_{r=1}^t\{\tilde{D}_{t,2(r-1)}-D_{t,2(r-1)}+\tilde{D}_{t,2r-1}\hat{H}-D_{t,2r-1}H\}(\hat\beta_{t-r}-\beta_{t-r})\\
&\quad+\sum_{r=1}^t(\tilde{D}_{t,2r-1}-D_{t,2r-1})\{(\beta_{t-r}^\top,-1)\otimes I_p\}\zeta.
\end{align*}
Thus, when $t\le T$, by \ref{cond:support}, 
\begin{equation}\label{eq:prelim}
\begin{aligned}
\normvec{\hat\beta_t-\beta_t-\Pi_t\zeta}
&\le\sum_{r=1}^t\normop{\{D_{t,2(r-1)}+D_{t,2r-1}H\}Q}\normvec{\hat\beta_{t-r}-\beta_{t-r}-\Pi_{t-r}\zeta}\\
&\quad+\sum_{r=1}^t\big[\normop{\{\tilde{D}_{t,2(r-1)}-D_{t,2(r-1)}\}Q}
+\normop{D_{t,2r-1}}\normop{(\hat{H}-H)Q}\\
&\hspace{4em}+\normop{\tilde{D}_{t,2r-1}-D_{t,2r-1}}\normop{\hat{H} Q}\big]
\normvec{\hat\beta_{t-r}-\beta_{t-r}}\\
&\quad+\sum_{r=1}^t\normop{\tilde{D}_{t,2r-1}-D_{t,2r-1}}\normop{(\beta_{t-r}^\top,-1)\otimes I_p}\normvec{\zeta}.
\end{aligned}
\end{equation}
Note that $\normop{(\beta^\top,-1)\otimes I_p}=\normsum{\beta}+1$,
$\normop{(\hat{H}-H)Q}\le s\normvec{\zeta}$ and $\normop{\hat{H} Q}\le Z$.
It follows that for all $t=0,1,\dots,T-1$, by \ref{cond:support}, 
\[\normvec{\hat\gamma_t-\gamma_t}\le Z\normvec{\hat\beta_t-\beta_t}+(\normsum{\beta_t}+1)\normvec{\zeta}.\]
If $t\le T$ and $\tilde\theta_{t-1}$ lies on the line segment connecting $\hat\theta_{t-1}$ and $\theta_{t-1}$, then 
\begin{align*}
\normvec{\tilde\theta_{t-1}-\theta_{t-1}}
&\le\normvec{\hat\theta_{t-1}-\theta_{t-1}}\\
&=\max_{r\le t-1}\max(\normvec{\hat\beta_r-\beta_r},\normvec{\hat\gamma_r-\gamma_r})\\
&\le\max_{r\le t-1}\{\max(Z,1)\normvec{\hat\beta_r-\beta_r}+(\normsum{\beta_r}+1)\normvec{\zeta}\}.
\end{align*}
We prove the target result together with $\normvec{\hat\beta_t-\beta_t}\le\delta_t$ by induction on $t$.
The case $t=0$ is trivially true. 
Suppose that the inductive hypothesis holds up to $t-1$. 
Then $\normvec{\hat\theta_{t-1}-\theta_{t-1}}\le\epsilon$ by \ref{cond:small}, so \eqref{eq:prelim} can be bounded as desired. 
Moreover, 
\begin{align*}
\normvec{\hat\beta_t-\beta_t}
&\le\normvec{\Pi_t\zeta}+a_t(\delta_{t-1}+\max_{r\le t-1}\normvec{\Pi_r\zeta})\\
&\quad+\omega_{t,1}(\epsilon)(\max_{r\le t-1}\normsum{\beta_r}+1)\normvec{\zeta}\\
&\quad+\{(\omega_{t,0}+Z\omega_{t,1})(\epsilon)+b_ts\normvec{\zeta}\big\}\delta_{t-1}\\
&\le\delta_t.
\end{align*}
This completes the proof.

\subsection*{Proof of Theorem~\ref{thm:linear_boot}}
Note that 
\[\hat\gamma_t^*-\hat\gamma_t
=\hat{H}^*(\hat\beta_t^*-\hat\beta_t)+(\hat{H}^*-\hat{H})\hat\beta_t-(\hat\xi^*-\hat\xi)
=\hat{H}^*(\hat\beta_t^*-\hat\beta_t)+\{(\hat\beta_t^\top,-1)\otimes I_p\}\zeta^*.\]
Then \eqref{eq:coef_boot} can be rewritten as 
\begin{align*}
\hat\beta_t^*-\hat\beta_t
&=\sum_{r=1}^t\big\{\hat{D}_{t,2(r-1)}(\hat\beta_{t-r}^*-\hat\beta_{t-r})+\hat{D}_{t,2r-1}(\hat\gamma_{t-r}^*-\hat\gamma_{t-r})\big\}\\
&=\sum_{r=1}^t\{\hat{D}_{t,2(r-1)}+\hat{D}_{t,2r-1}\hat{H}^*\}(\hat\beta_{t-r}^*-\hat\beta_{t-r})+\sum_{r=1}^t\hat{D}_{t,2r-1}\{(\hat\beta_{t-r}^\top,-1)\otimes I_p\}\zeta^*.
\end{align*}
Recalling \eqref{eq:sensitivity},
\begin{align*}
&\hat\beta_t^*-\hat\beta_t-\Pi_t\zeta^*\\
&=\sum_{r=1}^t\{D_{t,2(r-1)}+D_{t,2r-1}H\}(\hat\beta_{t-r}^*-\hat\beta_{t-r}-\Pi_{t-r}\zeta^*)\\
&\quad+\sum_{r=1}^t\{\hat{D}_{t,2(r-1)}-D_{t,2(r-1)}+\hat{D}_{t,2r-1}\hat{H}^*-D_{t,2r-1}H\}(\hat\beta_{t-r}^*-\hat\beta_{t-r})\\
&\quad+\sum_{r=1}^t\big[(\hat{D}_{t,2r-1}-D_{t,2r-1})\{(\beta_{t-r}^\top,-1)\otimes I_p\}+\hat{D}_{t,2r-1}\{(\hat\beta_{t-r}^\top-\beta_{t-r}^\top,0)\otimes I_p\}\big]\zeta^*.
\end{align*}
Thus, for all $t\le T$, 
\begin{equation}\label{eq:prelim_boot}
\begin{aligned}
\normvec{\hat\beta_t^*-\hat\beta_t-\Pi_t\zeta^*}
&\le\sum_{r=1}^t\normop{\{D_{t,2(r-1)}+D_{t,2r-1}H\}Q}\normvec{\hat\beta_{t-r}^*-\hat\beta_{t-r}-\Pi_{t-r}\zeta^*}\\
&\quad+\sum_{r=1}^t\big[\normop{\{\hat{D}_{t,2(r-1)}-D_{t,2(r-1)}\}Q}
+\normop{D_{t,2r-1}}\normop{(\hat{H}^*-H)Q}\\
&\hspace{4em}+\normop{\hat{D}_{t,2r-1}-D_{t,2r-1}}\normop{\hat{H}^*Q}\big]
\normvec{\hat\beta_{t-r}^*-\hat\beta_{t-r}}\\
&\quad+\sum_{r=1}^t\big[\normop{\hat{D}_{t,2r-1}-D_{t,2r-1}}\normop{(\beta_{t-r}^\top,-1)\otimes I_p}\\
&\hspace{4em}+\normop{\hat{D}_{t,2r-1}}\normop{(\hat\beta_{t-r}^\top-\beta_{t-r}^\top,0)\otimes I_p}\big]\normvec{\zeta^*}\\
&\le a_t\max_{r\le t-1}\normvec{\hat\beta_r^*-\hat\beta_r-\Pi_r\zeta^*}\\
&\quad+\{(\omega_{t,0}+Z^*\omega_{t,1})(\epsilon)+b_ts(\normvec{\zeta}+\normvec{\zeta^*})\}\max_{r\le t-1}\normvec{\hat\beta_r^*-\hat\beta_r}\\
&\quad+[\omega_{t,1}(\epsilon)(\max_{r\le t-1}\normsum{\beta_r}+1)+\{b_t+\omega_{t,1}(\epsilon)\}s\delta_{t-1}]\normvec{\zeta^*}.
\end{aligned}
\end{equation}
It follows that 
\begin{align*}
\normvec{\hat\beta_t^*-\hat\beta_t}
&\le\normvec{\Pi_t\zeta^*}+a_t\max_{r\le t-1}\normvec{\Pi_r\zeta^*}\\
&\quad+\{a_t+b_ts(\normvec{\zeta}+\normvec{\zeta^*})+(\omega_{t,0}+Z^*\omega_{t,1})(\epsilon)\}\max_{r\le t-1}\normvec{\hat\beta_r^*-\hat\beta_r}\\
&\quad+[\omega_{t,1}(\epsilon)(\max_{r\le t-1}\normsum{\beta_r}+1)+\{b_t+\omega_{t,1}(\epsilon)\}s\delta_{t-1}]\normvec{\zeta^*}.
\end{align*}
Induction on $t$ leads to $\normvec{\hat\beta_t^*-\hat\beta_t}\le\delta_t^*$, $t\le T$.
Plugging this back into \eqref{eq:prelim_boot} completes the proof.

\subsection*{Proof of Theorem~\ref{thm:valid}}
The distributional conclusion follows from Lemmas~\ref{lem:Gauss} and~\ref{lem:bootstrap} by the triangle inequality.
Write $F_n^G(x)=\Pr(S^G\le x)$ and $q_n(v)=\inf\{x:F_n^G(x)\ge v\}$ for $0<v<1$.
The anti-concentration inequality~\eqref{eq:cck-anti} implies that $F_n^G$ is continuous. Hence $F_n^G\{q_n(v)\}=v$.
Choose a deterministic $\eta_n\downarrow0$ such that
\[
\Pr\left\{\sup_x\abs{\Pr_{|W_1^n}(S_n^*\le x)-F_n^G(x)}>\eta_n\right\}\to0.
\]
On the complementary event, for all sufficiently large $n$,
\[
q_n(1-\alpha-2\eta_n)\le c_{1-\alpha}^*\le q_n(1-\alpha+2\eta_n).
\]
Applying Lemma~\ref{lem:Gauss} at these deterministic bounding quantiles gives
\[
1-\alpha-2\eta_n-o(1)\le\Pr(S_n\le c_{1-\alpha}^*)\le1-\alpha+2\eta_n+o(1).
\]
This proves the quantile claim and completes the proof.

\section{Proofs of Lemmas}
\subsection*{Proof of Lemma~\ref{lem:Gauss}}
For each $n$, define
\[
\mathcal{A}_n=\{\Pi_t^\top u:(u,t)\in\mathcal{U}\times\mathcal{T},\ \Pi_t^\top u\ne0_d\}.
\]
Choose a countable dense subset $\mathcal{A}_{n,0}\subset\mathcal{A}_n$ and set $\mathcal{F}_{n,0}=\{a^\top\ell:a\in\mathcal{A}_{n,0}\}$.
For every $a\in\mathcal{A}_n$, there are $a_m\in\mathcal{A}_{n,0}$ such that
\[
a_m\to a
\implies
a_m^\top\ell(w)\to a^\top\ell(w),\qquad w\in\operatorname{dom}(\ell).
\]
Thus, $\mathcal{F}_n$ is pointwise measurable.
Since $\mathcal{A}_n$ is nonempty and symmetric, for every $z\in\R^d$,
\[
\sup_{u,t}u^\top\Pi_tz
=\sup_{a\in\mathcal{A}_n}a^\top z
=\sup_{a\in\mathcal{A}_{n,0}}a^\top z\ge0.
\]
A countable dense subset of $\mathcal{U}\subset\R^p$, with $\mathcal{T}\subset\bN$, also gives measurable versions of $S_n$ and $S_n^*$.

Write $f_a=a^\top\ell$ and $\rho(f_a,f_b)=\{P_W(f_a-f_b)^2\}^{1/2}$.
The process $G_Pf_a=a^\top\Sigma^{1/2}g$ satisfies
\[
\begin{aligned}
\E(G_Pf_a\,G_Pf_b)&=a^\top\Sigma b=P_W(f_af_b),\\
|G_Pf_a-G_Pf_b|&\le\rho(f_a,f_b)\|g\|_2,\\
\sup_{a\in\mathcal{A}_n}|G_Pf_a|&\le\overline\sigma\|g\|_2.
\end{aligned}
\]
It is a tight Gaussian random element of $\ell^\infty(\mathcal{F}_n)$ with uniformly $\rho$-continuous paths, and its supremum equals $S^G$.

We first transfer~\eqref{eq:cck-entropy} to $P_W$.
Fix $0<r\le2\overline\sigma$ and any finite collection $f_1,\ldots,f_m$ with $\min_{j\ne k}\rho(f_j,f_k)>r/2$.
Finite empirical approximations to $P_W$ give a finitely supported $Q$ for which
\[
\min_{j\ne k}\|f_j-f_k\|_{L^2(Q)}>r/4,
\qquad
\|\bar f_n\|_{L^2(Q)}\le2B_n.
\]
Consequently, a ball of radius $(r/16B_n)\|\bar f_n\|_{L^2(Q)}$ contains at most one of the $f_j$, and
\[
m\le N\{(r/16B_n)\|\bar f_n\|_{L^2(Q)},\mathcal{F}_n,L^2(Q)\}
\le(16A_nB_n/r)^{v_n}.
\]
A maximal $(r/2)$-separated collection therefore gives
\[
N\{r,\mathcal{F}_n,L^2(P_W)\}\le(16A_nB_n/r)^{v_n}.
\]
The Gaussian entropy integral now yields
\begin{align*}
\E(S^G)
&\le C\int_0^{2\overline\sigma}\{v_n\log(16A_nB_n/r)\}^{1/2}\dd{r}\\
&\le C\overline\sigma v_n^{1/2}\left\{\log^{1/2}(8A_nB_n/\overline\sigma)+\int_0^1\log^{1/2}(1/x)\dd{x}\right\}\\
&\le CK_n^{1/2}.
\end{align*}
By \citet[Lemma~A.1]{chernozhukov2014suprema} and~\eqref{eq:cck-moments},
\begin{equation}\label{eq:cck-anti}
\omega_n^G(h):=\sup_x\Pr(|S^G-x|\le h)
\le Ch[K_n^{1/2}+\max\{\log^{1/2}(1/h),1\}],\qquad 0<h<1.
\end{equation}
In particular,
\begin{equation}\label{eq:cck-anti-scales}
\begin{aligned}
h_nK_n^{1/2}\to0&\implies\omega_n^G(h_n)\to0,\\
\limsup_{n\to\infty}\omega_n^G(\varepsilon/K_n^{1/2})&\le C\varepsilon,\qquad \varepsilon>0.
\end{aligned}
\end{equation}
The first implication uses $h_n\log^{1/2}(1/h_n)\to0$, and the second uses $K_n\ge\log n\to\infty$.

Let
\[
Z_n=\sup_{f\in\mathcal{F}_n}n^{-1/2}\sum_{i=1}^nf(W_i),\qquad
r_n=\frac{B_nK_n^{3/2}}{n^{1/2-1/q}}+\frac{B_n^{1/3}K_n^{7/6}}{n^{1/6}}.
\]
Condition~\eqref{eq:moment-cond} gives $r_n\to0$ and
\[
\frac{K_n^3}{n}\le\frac{B_n^2K_n^7}{n}\le r_n^6\to0.
\]
Apply \citet[Theorem~2.1]{chernozhukov2016couplings} with drift zero, $b=B_n$, $\sigma=\overline\sigma$, and the entropy constants in~\eqref{eq:cck-entropy}.
Its drift covering number is one, so its complexity parameter equals $K_n$.
For all sufficiently large $n$, choose
\[
\gamma_n=r_n^{1/4}\in(0,1),\qquad \eta_n=(nK_n^{1/2})^{-1}.
\]
On an extension of the probability space, the theorem provides $\widetilde Z_n\stackrel{d}{=}S^G$ such that
\[
\Pr(|Z_n-\widetilde Z_n|>d_{n,1})\le p_{n,1}:=C'(\gamma_n+n^{-1}),
\]
where
\[
d_{n,1}=C\left\{\eta_n+\frac{B_nK_n}{\gamma_n^{1/q}n^{1/2-1/q}}+\frac{(B_n\overline\sigma^2K_n^2)^{1/3}}{\gamma_n^{1/3}n^{1/6}}\right\}.
\]
Since $q\ge4$,
\[
K_n^{1/2}\,d_{n,1}
\le C(n^{-1}+r_n\gamma_n^{-1/q}+r_n\gamma_n^{-1/3})
\le C(n^{-1}+r_n^{11/12})\to0.
\]
Writing $F_n^G(x)=\Pr(S^G\le x)$, the coupling gives, for every $x\in\R$,
\[
F_n^G(x-d_{n,1})-p_{n,1}\le\Pr(Z_n\le x)\le F_n^G(x+d_{n,1})+p_{n,1}.
\]
Hence,
\begin{equation*}\label{eq:cck-leading}
\Delta_n^Z:=\sup_x|\Pr(Z_n\le x)-F_n^G(x)|
\le p_{n,1}+\omega_n^G(d_{n,1})\to0.
\end{equation*}

The input representation of $\zeta$ implies
\begin{align*}
|S_n-Z_n|
\le\sqrt{n}\sup_{u,t}|u^\top(\hat\beta_t-\beta_t-\Pi_t\zeta)|+\sqrt{n}\sup_{u,t}|u^\top\Pi_tR_n|
=o_{\Pr}(K_n^{-1/2}).
\end{align*}
For every fixed $\varepsilon>0$, put $h=\varepsilon/K_n^{1/2}$ and $p_n(h)=\Pr(|S_n-Z_n|>h)$.
Then
\[
F_n^G(x-h)-\Delta_n^Z-p_n(h)
\le\Pr(S_n\le x)
\le F_n^G(x+h)+\Delta_n^Z+p_n(h),
\]
so
\[
\limsup_{n\to\infty}\sup_x|\Pr(S_n\le x)-F_n^G(x)|
\le\limsup_{n\to\infty}\{\Delta_n^Z+p_n(h)+\omega_n^G(h)\}
\le C\varepsilon.
\]
Letting $\varepsilon\downarrow0$ completes the proof.

\subsection*{Proof of Lemma~\ref{lem:bootstrap}}
Consider the multiplier statistic formed from the centered true influence vectors,
\[
Z_n^e=\sup_{u,t}u^\top\Pi_t\Big[n^{-1/2}\sum_{i=1}^n\varpi_i\{\ell(W_i)-\bar\ell\}\Big]
=\sup_{f\in\mathcal{F}_n}n^{-1/2}\sum_{i=1}^n\varpi_i\{f(W_i)-P_nf\},
\]
where $P_nf=n^{-1}\sum_i f(W_i)$.
Use $r_n$, $\gamma_n$ and $\eta_n$ from the proof of Lemma~\ref{lem:Gauss}.
Since $K_n^3/n\to0$, the condition $K_n\le n$ in \citet[Theorem~2.2]{chernozhukov2016couplings} holds for all sufficiently large $n$.
That theorem, with drift zero, gives a coupled variable $\widetilde Z_n^e$ such that
\[
\Pr_{|W_1^n}(\widetilde Z_n^e\le x)=F_n^G(x),\qquad x\in\R,
\]
and
\[
\Pr(|Z_n^e-\widetilde Z_n^e|>d_{n,2})\le C'(\gamma_n+n^{-1}),
\]
where
\[
d_{n,2}=C\left\{\eta_n+\frac{B_nK_n}{\gamma_n^{1+1/q}n^{1/2-1/q}}+\frac{(B_n\overline\sigma K_n^{3/2})^{1/2}}{\gamma_n^{1+1/q}n^{1/4}}\right\}.
\]
The second component satisfies
\[
\left\{K_n^{1/2}\,\frac{(B_n\overline\sigma K_n^{3/2})^{1/2}}{n^{1/4}}\right\}^4
=\frac{B_n^2\overline\sigma^2K_n^5}{n}
\le\overline\sigma^2r_n^6.
\]
Therefore,
\begin{align*}
K_n^{1/2}\,d_{n,2}
\le C\{n^{-1}+\gamma_n^{-1-1/q}(r_n+r_n^{3/2})\}
\le C\{n^{-1}+r_n^{1-(1+1/q)/4}\}\to0.
\end{align*}
Define $p_n^e=\Pr_{|W_1^n}(|Z_n^e-\widetilde Z_n^e|>d_{n,2})$.
For every $\delta>0$,
\[
\Pr(p_n^e>\delta)
\le\delta^{-1}\E(p_n^e)
\le\frac{C'(\gamma_n+n^{-1})}{\delta}\to0.
\]
The conditional coupling yields
\[
F_n^G(x-d_{n,2})-p_n^e
\le\Pr_{|W_1^n}(Z_n^e\le x)
\le F_n^G(x+d_{n,2})+p_n^e,
\]
and hence, by~\eqref{eq:cck-anti-scales},
\begin{equation*}\label{eq:cck-multiplier}
\Delta_n^e:=\sup_x|\Pr_{|W_1^n}(Z_n^e\le x)-F_n^G(x)|
\le p_n^e+\omega_n^G(d_{n,2})=o_{\Pr}(1).
\end{equation*}

To compare the estimated and true influence vectors, set
\[
E_i=\hat\ell_i-\ell(W_i)+\bar\ell,\qquad
V_n=n^{-1/2}\sum_{i=1}^n\varpi_iE_i,\qquad
\widehat Z_n^e=\sup_{u,t}\sqrt{n}\,u^\top\Pi_t\zeta^*.
\]
We have 
\[
|\widehat Z_n^e-Z_n^e|
\le\sup_{u,t}|u^\top\Pi_tV_n|
\le J_n\|V_n\|_\infty.
\]
Conditionally on the data, $V_{n,j}$ is centered Gaussian and
\[
\Var_{|W_1^n}(V_{n,j})=n^{-1}\sum_iE_{ij}^2\le e_n^2.
\]
For every $\theta>0$,
\[
\E_{|W_1^n}\{\exp(\theta\|V_n\|_\infty)\}
\le\sum_{j=1}^d\sum_{s\in\{-1,1\}}\E_{|W_1^n}\{\exp(s\theta V_{n,j})\}
\le2d\exp(\theta^2e_n^2/2).
\]
Jensen's inequality and minimization over $\theta$ imply
\[
\E_{|W_1^n}\{\|V_n\|_\infty\}
\le\inf_{\theta>0}\left\{\frac{\log(2d)}{\theta}+\frac{\theta e_n^2}{2}\right\}
=e_n\sqrt{2\log(2d)}.
\]
Thus, for every $\varepsilon>0$,
\[
\Pr_{|W_1^n}\{K_n^{1/2}|\widehat Z_n^e-Z_n^e|>\varepsilon\}
\le\frac{J_ne_n\sqrt{2K_n\log(2d)}}{\varepsilon}
=o_{\Pr}(1).
\]
Condition~\eqref{eq:linear_boot-cond} also gives
\[
|S_n^*-\widehat Z_n^e|
\le\sqrt{n}\sup_{u,t}|u^\top(\hat\beta_t^*-\hat\beta_t-\Pi_t\zeta^*)|
=o_{\Pr_{|W_1^n}}(K_n^{-1/2})
\]
in probability.
For $h=\varepsilon/K_n^{1/2}$, it follows that
\begin{align*}
p_n^*(h)&:=\Pr_{|W_1^n}(|S_n^*-Z_n^e|>h)\\
&\le\Pr_{|W_1^n}(|S_n^*-\widehat Z_n^e|>h/2)+\Pr_{|W_1^n}(|\widehat Z_n^e-Z_n^e|>h/2)
=o_{\Pr}(1).
\end{align*}
For every $x\in\R$,
\[
F_n^G(x-h)-\Delta_n^e-p_n^*(h)
\le\Pr_{|W_1^n}(S_n^*\le x)
\le F_n^G(x+h)+\Delta_n^e+p_n^*(h).
\]
Consequently,
\[
\sup_x|\Pr_{|W_1^n}(S_n^*\le x)-F_n^G(x)|
\le\Delta_n^e+p_n^*(h)+\omega_n^G(h).
\]
Given $\delta>0$, choose $\varepsilon>0$ so that $C\varepsilon<\delta/4$ in~\eqref{eq:cck-anti-scales}.
For all sufficiently large $n$, $\omega_n^G(h)<\delta/2$, whence
\[
\Pr\left\{\sup_x|\Pr_{|W_1^n}(S_n^*\le x)-F_n^G(x)|>\delta\right\}
\le\Pr\{\Delta_n^e+p_n^*(h)>\delta/2\}\to0.
\]
This completes the proof.

\section{Verifying the conditions in Lemmas}
Let \[\mathcal{A}_n=\{\Pi_t^\top u:(u,t)\in\mathcal{U}\times\mathcal{T},\ \Pi_t^\top u\ne0_d\}\] and suppose $J_n=\sup_{a\in\mathcal{A}_n}\|a\|_1<\infty$.
For the envelope $\bar f_n(w)=J_n\|\ell(w)\|_\infty$,
\[
\|(a-b)^\top\ell\|_{L^2(Q)}
\le\|a-b\|_1\,\big\|\|\ell\|_\infty\big\|_{L^2(Q)}
=\frac{\|a-b\|_1}{J_n}\|\bar f_n\|_{L^2(Q)}.
\]
Thus, for every finitely supported $Q$,
\begin{equation*}\label{eq:cck-coefficient-cover}
N\{\varepsilon\|\bar f_n\|_{L^2(Q)},\mathcal{F}_n,L^2(Q)\}
\le N(\varepsilon J_n,\mathcal{A}_n,\|\cdot\|_1).
\end{equation*}
Suppose $\mathcal{A}_n$ lies in a linear subspace $\mathcal{V}_n$ of dimension $k_n\ge1$.
The balls of radius $\varepsilon J_n/2$ about an $(\varepsilon J_n)$-separated collection in $\mathcal{A}_n$ are disjoint and contained in the ball of radius $(1+\varepsilon/2)J_n$ in $\mathcal{V}_n$.
Comparing their $k_n$-dimensional volumes gives
\[
N(\varepsilon J_n,\mathcal{A}_n,\|\cdot\|_1)
\le(1+2/\varepsilon)^{k_n}
\le(5/\varepsilon)^{k_n},\qquad 0<\varepsilon\le1.
\]
Hence $A_n=5$ and $v_n=k_n$ satisfy~\eqref{eq:cck-entropy}.
If instead $\|a\|_0\le k_n\le d$ for all $a\in\mathcal{A}_n$, cover the subsets supported on each set of $k_n$ coordinates to obtain
\[
N(\varepsilon J_n,\mathcal{A}_n,\|\cdot\|_1)
\le\binom{d}{k_n}(5/\varepsilon)^{k_n}
\le\left(\frac{5\mathrm e d}{k_n\varepsilon}\right)^{k_n}.
\]
In this case, $A_n=5\mathrm e d/k_n$ and $v_n=k_n$ are admissible.

The variance condition is
\[
\underline\sigma^2\le a^\top\Sigma a\le\overline\sigma^2,
\qquad a\in\mathcal{A}_n.
\]
For example, if $\mathcal{V}_n=\operatorname{span}(\mathcal{A}_n)$ and constants $c_0,c_1>0$ satisfy
\[
\inf_{a\in\mathcal{A}_n}\|a\|_2\ge c_0,
\qquad
\inf_{v\in \mathcal{V}_n:\|v\|_2=1}v^\top\Sigma v\ge c_1,
\]
then $\inf_{a\in\mathcal{A}_n}a^\top\Sigma a\ge c_0^2c_1$.
When a contrast set is a norm ball, an equivalent norming set can be used; for instance,
\[
\sup_{\|u\|_1\le1}u^\top z
=\|z\|_\infty
=\max_{u\in\{\pm e_1,\ldots,\pm e_p\}}u^\top z,
\qquad z\in\R^p.
\]

Let $M_{n,k}=\sup_{a\in\mathcal{A}_n}P_W|a^\top\ell|^k$.
Given $M_{n,2}\le\overline\sigma^2$, the upper bounds in~\eqref{eq:cck-moments} hold whenever
\[
B_n\ge\max\Big\{1,\overline\sigma,\ J_n\big\|\|\ell(W)\|_\infty\big\|_{L^q},\ M_{n,3}/\overline\sigma^2,M_{n,4}^{1/2}/\overline\sigma\Big\}.
\]
For exact sample-average inputs, let $h(W_i)\in\R^d$ stack their matrix and vector contributions, and put $\vartheta=P_Wh$ and $\hat\vartheta=P_nh$.
Then
\[
\begin{aligned}
\zeta&=\hat\vartheta-\vartheta=P_n\ell,\qquad \ell=h-\vartheta,\qquad R_n=0_d,\\
\hat\ell_i&=h(W_i)-\hat\vartheta=\ell(W_i)-P_n\ell,\qquad e_n=0.
\end{aligned}
\]

We next derive uniform bounds from the trajectory recursions.
Work on the neighborhood and support events of Theorems~\ref{thm:linear} and~\ref{thm:linear_boot} for every finite horizon, and assume
\[
\sup_{t\ge1}a_t\le a<1,\qquad
b:=\sup_{t\ge1}b_t<\infty,\qquad
M:=\sup_{t\ge0}\|\beta_t\|_1<\infty.
\]
For the chosen radius $\epsilon_n$, write
\[
\begin{gathered}
\Omega_j=\sup_{t\ge1}\omega_{t,j}(\epsilon_n),\quad j=0,1,\qquad
h=\|\zeta\|_\infty,\qquad h^*=\|\zeta^*\|_\infty,\\
L=\sup_{t\ge0}\|\Pi_t\zeta\|_\infty,\qquad
L^*=\sup_{t\ge0}\|\Pi_t\zeta^*\|_\infty.
\end{gathered}
\]
Set
\[
\mathsf{A}=a+bsh+\Omega_0+Z\Omega_1,\qquad
\mathsf{B}=(1+a)L+\Omega_1(M+1)h.
\]
If $\mathsf{A}<1$, the recursion for $\delta_t$ gives
\[
\delta_0=0,\qquad
\delta_t\le\max(\delta_{t-1},\mathsf{A}\delta_{t-1}+\mathsf{B})
\implies
\Delta:=\sup_t\delta_t\le\frac{\mathsf{B}}{1-\mathsf{A}}.
\]
For the bootstrap, define
\[
\begin{aligned}
\mathsf{A}^*&=a+bs(h+h^*)+\Omega_0+Z^*\Omega_1,\\
\mathsf{B}^*&=(1+a)L^*+\{\Omega_1(M+1)+(b+\Omega_1)s\Delta\}h^*.
\end{aligned}
\]
The same induction, when $\mathsf{A}^*<1$, yields
\[
\delta_0^*=0,\qquad
\delta_t^*\le\max(\delta_{t-1}^*,\mathsf{A}^*\delta_{t-1}^*+\mathsf{B}^*)
\implies
\Delta^*:=\sup_t\delta_t^*\le\frac{\mathsf{B}^*}{1-\mathsf{A}^*}.
\]
Let $r_t=\|\hat\beta_t-\beta_t-\Pi_t\zeta\|_\infty$ and $r_t^*=\|\hat\beta_t^*-\hat\beta_t-\Pi_t\zeta^*\|_\infty$.
The two trajectory bounds imply
\[
\begin{aligned}
r_t&\le a\max_{k<t}r_k+R,\\
R&=\Omega_1(M+1)h+(bsh+\Omega_0+Z\Omega_1)\Delta,\\
r_t^*&\le a\max_{k<t}r_k^*+R^*,\\
R^*&=\{bs(h+h^*)+\Omega_0+Z^*\Omega_1\}\Delta^*
+\{\Omega_1(M+1)+(b+\Omega_1)s\Delta\}h^*.
\end{aligned}
\]
Since $r_0=r_0^*=0$, induction gives, for every finite $T$,
\[
\max_{t\le T}r_t\le\frac{R}{1-a},\qquad
\max_{t\le T}r_t^*\le\frac{R^*}{1-a}.
\]
Letting $T\to\infty$ and putting $U_n=\sup_{u\in\mathcal{U}}\|u\|_1$ yields
\begin{equation}\label{eq:cck-uniform-remainders}
\begin{aligned}
\sup_{u,t}|u^\top(\hat\beta_t-\beta_t-\Pi_t\zeta)|&\le\frac{U_nR}{1-a},\\
\sup_{u,t}|u^\top(\hat\beta_t^*-\hat\beta_t-\Pi_t\zeta^*)|&\le\frac{U_nR^*}{1-a}.
\end{aligned}
\end{equation}
These bounds hold on the full-history events; assume their sampling probability tends to one and their conditional bootstrap probability tends to one in probability.
If $a<1$ is fixed, $b,M,Z,Z^*$ are bounded in the respective probability senses, and, for a deterministic $\nu_n\to0$,
\[
\begin{gathered}
h+L+\Omega_0+\Omega_1=\mathcal{O}_{\Pr}(\nu_n),\qquad
h^*+L^*=\mathcal{O}_{\Pr_{|W_1^n}}(\nu_n)\ \text{in probability},\qquad s\nu_n\to0,
\end{gathered}
\]
then the preceding formulas give
\[
\begin{aligned}
\mathsf{A}-a&=\mathcal{O}_{\Pr}(s\nu_n),&
\mathsf{B}&=\mathcal{O}_{\Pr}(\nu_n),&
\Delta&=\mathcal{O}_{\Pr}(\nu_n),&
R&=\mathcal{O}_{\Pr}(s\nu_n^2),\\
\mathsf{A}^*-a&=\mathcal{O}_{\Pr_{|W_1^n}}(s\nu_n),&
\mathsf{B}^*&=\mathcal{O}_{\Pr_{|W_1^n}}(\nu_n),&
\Delta^*&=\mathcal{O}_{\Pr_{|W_1^n}}(\nu_n),&
R^*&=\mathcal{O}_{\Pr_{|W_1^n}}(s\nu_n^2),
\end{aligned}
\]
where the conditional orders hold in probability.
Consequently, sufficient rates for~\eqref{eq:linear-cond} and~\eqref{eq:linear_boot-cond} are
\[
\sqrt{nK_n}\,U_ns\nu_n^2\to0,\qquad
\sqrt{nK_n}\sup_{u,t}|u^\top\Pi_tR_n|=o_{\Pr}(1).
\]
For $U_n=\mathcal{O}(1)$ and $\nu_n=\{\log(\mathrm e d)/n\}^{1/2}$, the first rate becomes
\[
s\log(\mathrm e d)\sqrt{K_n/n}\to0.
\]

For soft thresholding, let $z_t$ denote the population pre-threshold vector in Example~\ref{eg:Jacobian}, and assume $\mu_t,\eta_t\ge0$.
If $\|\tilde\theta_{t-1}-\theta_{t-1}\|_\infty\le\epsilon_n$, then
\[
\begin{aligned}
\|\tilde z_t-z_t\|_\infty
&\le\{(1+\mu_t)+\mu_t\}(1+\eta_t)\epsilon_n\\
&=(1+2\mu_t)(1+\eta_t)\epsilon_n.
\end{aligned}
\]
Hence the uniform margin condition
\[
\inf_{t,j}\big||z_{t,j}|-\lambda\eta_t\big|
>\sup_t(1+2\mu_t)(1+\eta_t)\epsilon_n
\]
implies
\[
\1\{|\tilde z_{t,j}|>\lambda\eta_t\}
=\1\{|z_{t,j}|>\lambda\eta_t\}
\implies
\Omega_0=\Omega_1=0.
\]

\section{Additional numerical results}
\begin{figure}[!ht]
\centering
\includegraphics[width=\linewidth]{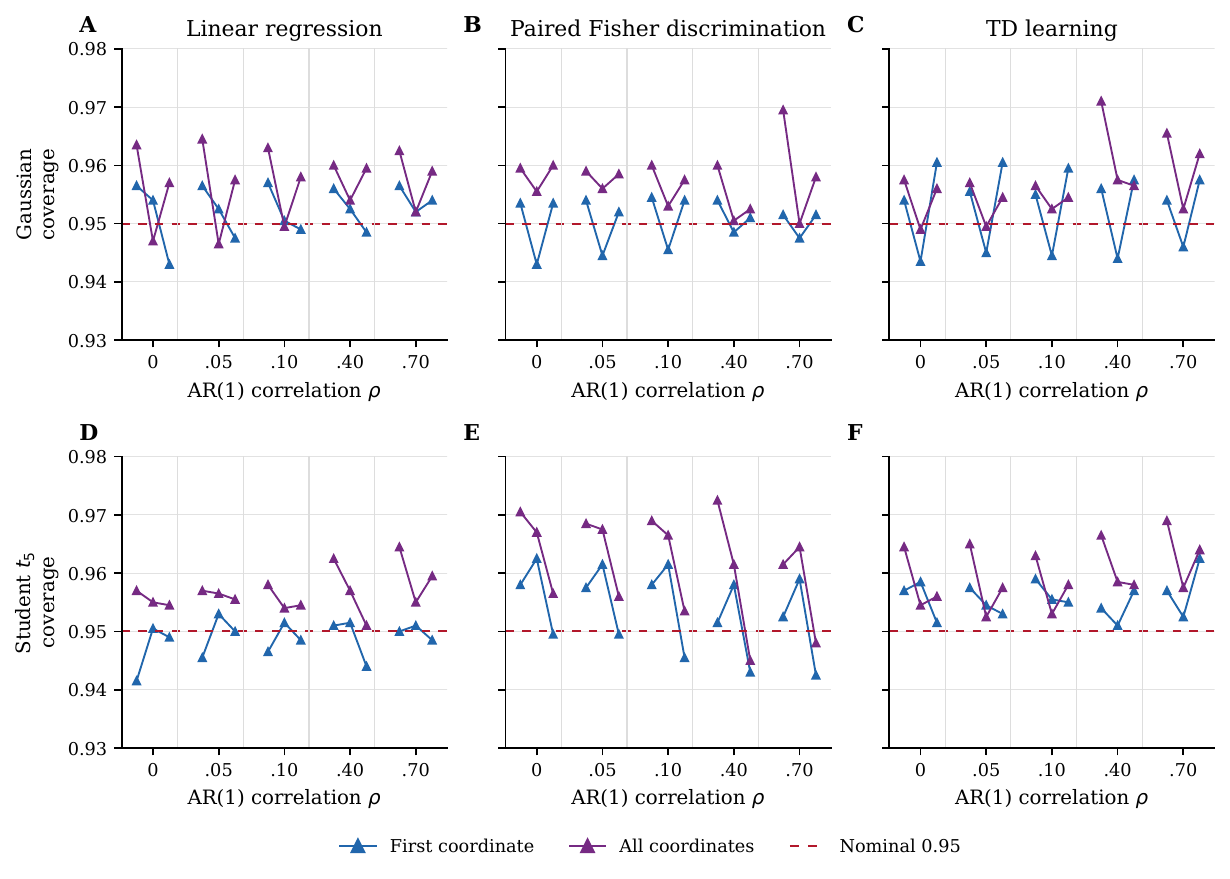}
\caption{Complete simultaneous-coverage results. 
Within each value of $\rho$, the three connected points are $n=300,600,1000$. 
The first row uses Gaussian noise and the second row uses standardized Student-$t_5$ disturbances.}
\label{fig:coverage-all}
\end{figure}

Figure~\ref{fig:coverage-all} reports all 180 coverage estimates: three estimating equations, two noise distributions, five correlation levels, three sample sizes, and two inferential targets. 
Within each correlation block, the three connected points correspond to $n=300,600,1000$ from left to right. 
The all-coordinate bands are mildly conservative more often than the first-coordinate bands. 
Coverage also remains close to nominal in many high-correlation configurations, but those configurations fail the sufficient population stability condition and are interpreted only as stress tests.

\begin{figure}[!ht]
\centering
\includegraphics[width=0.7\linewidth]{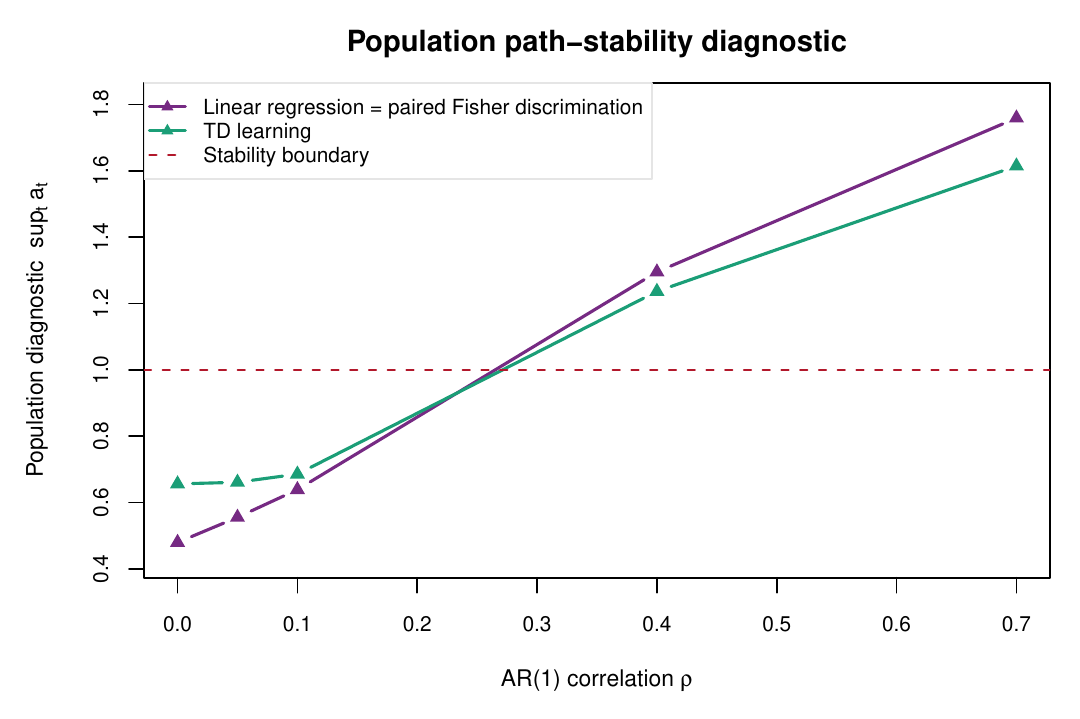}
\caption{Population path-stability diagnostic. 
The curves coincide for linear regression and paired Fisher discrimination, and are therefore represented by one line.
Configurations with $\rho\le0.10$ provide the primary evidence, whereas the
two higher-correlation settings are stress tests.}
\label{fig:stability}
\end{figure}

The population diagnostic in Figure~\ref{fig:stability} explains the reporting split used in the main paper. 
With the implemented step sizes, $\sup_ta_t<1$ for $\rho\le0.10$. 
At $\rho=0.40$ and $0.70$ it exceeds one in every equation. 

\clearpage
\phantomsection
\addcontentsline{toc}{section}{References}
\bibliographystyle{apalike}
\bibliography{ref}
\end{document}